\documentclass[english,nofootinbib]{revtex4-2}
\usepackage[T1]{fontenc}
\usepackage[latin9]{inputenc}
\usepackage{xcolor}
\usepackage{array}
\usepackage{float}
\usepackage{units}
\usepackage{mathrsfs}
\usepackage{mathtools}
\usepackage{amsmath}
\usepackage{stackrel}
\usepackage{graphicx}
\usepackage{dsfont}

\usepackage{epstopdf}
\makeatletter
\providecommand{\tabularnewline}{\\}

\makeatother

\usepackage{babel}
\begin{document}
\title{Collective attack free controlled quantum key agreement without quantum memory}
\author{Arindam Dutta}
\email{Corresponding author: arindamsalt@gmail.com}

\email{https://orcid.org/0000-0003-3909-7519}

\author{Anirban Pathak}
\email{anirban.pathak@gmail.com}

\email{https://orcid.org/0000-0003-4195-2588}

\affiliation{Department of Physics and Materials Science \& Engineering, Jaypee
Institute of Information Technology, A 10, Sector 62, Noida, UP-201309,
India}
\begin{abstract}
Here we present a new protocol for controlled quantum key agreement
and another protocol for key agreement with a specific focus on the
security analysis. Specifically, detailed security proof is provided
against impersonated fraudulent attack and collective attacks and
it is established that the proposed protocols are not only secure,
but they also satisfy other desired properties of such schemes (i.e.,
fairness and correctness). Further, the proposed schemes are critically
compared with a set of schemes for quantum key agreement and an existing
scheme for controlled quantum key agreement (Tang et al.'s protocol)
in terms of efficiency and the required quantum resources. Especially,
it is observed that in contrast to the existing schemes, the present
scheme does not require quantum memory. In addition, the protocol
for controlled quantum key agreement proposed here is found to require
quantum resources (Bell state and single photon state) that are easier
to produce and maintain compared to the quantum resources (GHZ states)
needed for the only known existing protocol for the same purpose,
i.e., Tang et al.'s protocol.
\end{abstract}
\maketitle

\section{Introduction\label{sec:Introduction}}

Key agreement (KA) is an extremely important research domain in cryptography
and is considered to be essential for achieving perfect forward secrecy.
A scheme for KA allows two or more parties to agree on a key (i.e.,
they jointly generate a random number) in a manner that all the legitimate
parties influence the outcome (final key) and neither a proper subset
of legitimate parties nor an eavesdropper can force a key on the legitimate
parties. This is a weaker definition of KA. A relatively stronger
definition to be followed here and usually in the literature of quantum
KA (QKA) demands that all legitimate users equally influence the final
key. In 1976, Diffie and Hellman first proposed a classical KA protocol
involving two parties \citep{DH76}, further efforts were done to
extend the Diffie-Hellman two-party scenario to multiparty settings
for more efficient KA protocols \citep{IT+82,MW+98,ST+2000,AS+2000}.
As mentioned above, if we follow the stronger definition of KA, each
legitimate party would equally contribute to the creation of an identical
shared key. This property that the key should be established with
equal influence by the legitimate participants and one or more participants
individually or collectively would not be able to influence the final
key is often referred to as ``fairness'' \citep{HWL+14,HSL+17}. 

As we know, security of the classical cryptography in general and
KA in particular arise from the computational complexity of a computational
task. With the steady development of quantum computers, the classical
notion of security is facing serious challenges as many of the computational
problems forming the backbone of the security of the schemes for classical
cryptography can be solved in polynomial time if scalable quantum
computers can be built. To conquer this challenge, various features
of quantum mechanics (e.g., collapse on measurement, no-cloning theorem,
Heisenberg's uncertainty principle, nonlocality, contextuality) are
exploited and it has been established that many of the cryptographic
tasks can be performed with unconditional security with the appropriate
use of quantum resources. One such task is KA whose quantum version
is QKA \cite{HJP20}. In this letter, we aim to propose two new schemes for QKA and
provide security proof for that. However, before moving to QKA and
its multipartite version, it will be apt to briefly mention about
the other aspects of quantum cryptography as that would clarify the
distinctive and unique nature of QKA and provide a brief idea about
other facets of quantum cryptography.

In 1984, Bennett and Brassard proposed the most celebrated BB84 protocol
\citep{BB84}, which magnificently achieves a key distribution task
between two remote parties in an ideal scenario using quantum resources.
The security of the BB84 protocol relies on the fundamental attributes
of quantum physics. Since the time BB84 protocol was proposed,
quantum key distribution (QKD) has evolved as a prominent and rapidly
advancing sub-field of cryptography, with significant progress in
both theory \citep{CPC+09,SPR17,P13} and experiments \citep{BRM+20,RRE+21,ZLX+23}.
Theoretical QKD schemes include those using single-photon \citep{B92,DP+23,SAR+04}
and entangled-photon sources \citep{E91,BBM92}. More recently, measurement
device-independent QKD \citep{LCQ12,CXC+14} and twin field QKD \citep{LYD+18,LYZ+19}
have gained popularity in the cryptographic community. Efforts have
also been made to enable free-space communication for QKD implementation
\citep{LCP+22,DBD+24,DMB+24} with specific attention on the satellite
assisted quantum communication \citep{GBL+23}. Furthermore, many
other important aspects of quantum cryptography, like quantum secure
direct communication (QSDC) \citep{ping-pong_BF_02,DL+03}, quantum
secret sharing (QSS) \citep{HB+99,KK+99,MSP+15,LLW+23}, quantum identity
authentication (QIA) \citep{DHHM_1999,DP22,DP23}, quantum private
comparison \citep{YW09,YC+09,CX+10,KSP18}, quantum digital signature
\citep{GC01,DW+14,WD+15}, have been developed. In addition to the
above, QKA has appeared as an important cryptographic primitive which
helps two or more parties to agree on an identical key with equal
contribution of their own keys by using quantum resources. The relevance
of QKA lies in its potential applications in electronic auctions,
multiparty secure computing, access control, etc. \citep{ST+2000},
and in the role that it plays in establishing forward secrecy\footnote{This security definition can be used to establish a secure bound for
QKA \citep{DP24}.}. Because of its relevance in different areas, QKA has recently been
established as a new domain of quantum cryptography. 

To the best of our knowledge, in 2004, Zhou et al. proposed the first
QKA scheme \citep{ZZX04}, in which two users use to apply quantum
teleportation to agree on an identical key. However, in 2009, Tsai
and Hwang \citep{TH09} pointed out that in Zhou et al.'s protocol,
one party can fully influence the shared key and then distribute that
key to another party without being detected. Thus, the protocol was
not fair. In fact, Zhou et al. protocol was not even secure and the
same was established in Ref. \citep{CT+11} in 2011. Two years later,
Shi and Zhou presented the first multiparty QKA (MQKA) protocol based
on entanglement swapping \citep{SZ13}. Unfortunately, Shi and Zhou's
scheme was also found to be insecure \citep{LG+13}. In between, Chong
et al. presented a QKA protocol based on the BB84 protocol, utilizing
delay measurement technology and an authenticated classical channel
\citep{CH10}. Later, several two-party QKA protocols were generalized
to MQKA protocols \citep{LG+13,SA+14,XW+14,HM15}, and these MQKA
protocols can be divided into three categories \citep{LX+16}; tree
type \citep{XW+14}, complete-graph-type \citep{LG+13}, and circle-type
(\citep{LZ+21} and Refs. therein). The last type is more feasible
to implement and has higher efficiency. In fact, quantum conference key agreement
(QCKA) and MQKA have become active areas of research \citep{MGK+20,PHU+23}.
These technologies utilize differential phase reference method \citep{IH24}
or discrete variable entangled quantum states. Recent experimental
progress in QCKA has been demonstrated by Proietti et al. \citep{PHG+21}.
Additionally, efforts to achieve mutually authenticated QKA using
new schemes have been explored by Wang et al. \citep{WZL24}.

A secure QKA protocol should satisfy the following three conditions;
\emph{Correctness:} After a successful run of the protocol, each participant
obtains the correct agreement key, \emph{Security: }No external (illegitimate)
party can eavesdrop to obtain any information about the final agreement
key without being detected, \emph{Fairness:} All participants equally
influence the final agreement key i.e., any proper subset of the participants
should not determine the agreement key alone. Utilizing these above
conditions, it is the first time that we propose a QKA protocol with
controller Charlie to accomplish the key agreement task between two
legitimate parties without employing quantum memory. Our scheme is
designed to address this kind of demand for quantum cryptography\footnote{We prove in Section \ref{sec:Security-Analysis} that our protocol
is secure against the impersonated fraudulent attack that also justifies
the third party can be considered as untrusted/dishonest.}. Such scenarios are common in quantum cryptography and several schemes
for controlled quantum cryptography have been proposed in the recent
past. A significant number of studies on controlled-QKD \citep{DGL+03, SNG+09}, controlled quantum dialogue \citep{DXG+08, KH17}, and controlled secure direct quantum communication \citep{WZT06, CWG+08, LLL+13, DP23}, among others, have been reported in recent years. However, to the best of our knowledge, no scheme for controlled-QKA (CQKA) \citep{TS+20} without relying on quantum memory has been proposed to date. Motivated by this fact, here we aim
to propose a scheme for CQKA. Here, we use Bell state, and single
qubit states as quantum resources to implement our CQKA protocol.
One-way quantum channel is used to circumvent unnecessary noise, this
is in contrast to a set of QKA protocols that need two-way quantum
channels \citep{SA+14,HM15,TS+20} leading to undesirable channel
noise. The impact of Markovian and non-Markovian channel on our protocol is examined,
and the fidelity is analyzed as a function of the noise parameter. In what follows, the security of the proposed scheme is proven
and it is established that the proposed CQKA scheme can be realized
using the currently available technology. Further, it is established
that the proposed scheme is more efficient in comparison with many
of the existing schemes. In addition, the presence of a controller
adds advantage in specific situations. 

The rest of this paper is organized as follows: In Section \ref{sec:New-QKA-protocol},
we explain the preliminary ideas required for describing our schemes
and subsequently describe our protocol in a step-wise manner. Then,
the security analysis against impersonated fraudulent attack, and
more general attacks (collective attack) is done in Section \ref{sec:Security-Analysis}.
The impact of noise on the protocol is discussed
in Section \ref{sec:Effect_of_Noise}. In Section \ref{sec:Comparison},
our protocol is critically compared with the existing protocols. A
discussion of the results and details of an experimental setup are
presented in Section \ref{sec:Discussion}. Finally, the work is concluded
in Section \ref{sec:Conclusion}.

\section{New CQKA protocol\label{sec:New-QKA-protocol}}

In the previous section, we have already mentioned that \emph{correctness
}is an important condition for a QKA protocol. In this section, we
elaborately describe our new CQKA protocol with specific attention
to the \emph{correctness }criteria. For this purpose, we take a specific
example or individual choice for $i^{th}$ key bit by Alice, Bob,
and Charlie to prepare their $i^{th}$ state of their qubit sequence
for further proceedings in the protocol. The first part of this section
may be considered as the description of steps taken to satisfy the
\emph{correctness }condition, and in the second part, we describe
our scheme in a conventional step-wise manner to elucidate in a more
general way.

\subsection*{Preliminary idea for proposed CQKA}

Here, we describe our new CQKA protocol based on the fundamental attributes
of quantum mechanics. To present the same, we need to state the relation
between the classical bit values with the quantum states. Here, we
use two types of relations (maps); one is for public announcement,
and the other one is for the final agreement key generation by Alice
and Bob \citep{SHW16}, which will not be announced publicly for satisfying
\emph{privacy} criteria. Charlie and Alice publicly announce classical
bit sequences $k_{C}$ and $k_{A}$ depending on the quantum states
produced by them during the execution of the protocol using the relations
mentioned in Eqs. (\ref{eq:Charlie's corresponding relation}) and
(\ref{eq:Alice's corresponding relation}), respectively. Bob announces
classical bit sequence $k_{B}$ determined by his measurement outcomes
in accordance with the mapping described through Eq. (\ref{eq:Bob's corresponding relation}).
The mapping of these classical sequences with their quantum counter
part is mentioned in the following: 

\begin{equation}
\begin{array}{lcl}
0 & : & |\phi^{+}\rangle=\frac{1}{\sqrt{2}}\left(\vert00\rangle+\vert11\rangle\right),\\
1 & : & \vert\phi^{-}\rangle=\frac{1}{\sqrt{2}}\left(\vert00\rangle-\vert11\rangle\right),
\end{array}\label{eq:Charlie's corresponding relation}
\end{equation}
which implies that if Charlie prepares $|\phi^{+}\rangle$ $(\vert\phi^{-}\rangle)$
then the corresponding bit value in the sequence $k_{C}$ will be
0 (1). Maps used by Alice and Bob described below can be visualized
in the same manner.

\begin{equation}
\begin{array}{lcl}
0 & : & |0\rangle,\\
1 & : & |1\rangle,
\end{array}\label{eq:Alice's corresponding relation}
\end{equation}
and
\begin{equation}
\begin{array}{ccccc}
0 & : & \vert\phi^{+}\rangle=\frac{1}{\sqrt{2}}\left(\vert00\rangle+\vert11\rangle\right) & {\rm or} & \vert\psi^{-}\rangle=\frac{1}{\sqrt{2}}\left(\vert01\rangle-\vert10\rangle\right),\\
1 & : & \vert\phi^{-}\rangle=\frac{1}{\sqrt{2}}\left(\vert00\rangle-\vert11\rangle\right) & {\rm or} & \vert\psi^{+}\rangle=\frac{1}{\sqrt{2}}\left(\vert01\rangle+\vert10\rangle\right).
\end{array}\label{eq:Bob's corresponding relation}
\end{equation}
Again, Alice and Bob use a mapping of two-bit classical information
with their final result after performing Bell measurement to get the
final agreement key. The mapping is in the following: 

\begin{equation}
\begin{array}{c}
00:\vert\phi^{+}\rangle=\frac{1}{\sqrt{2}}\left(\vert00\rangle+\vert11\rangle\right),\\
01:\vert\phi^{-}\rangle=\frac{1}{\sqrt{2}}\left(\vert00\rangle-\vert11\rangle\right),\\
10:\vert\psi^{+}\rangle=\frac{1}{\sqrt{2}}\left(\vert01\rangle+\vert10\rangle\right),\\
11:\vert\psi^{-}\rangle=\frac{1}{\sqrt{2}}\left(\vert01\rangle-\vert10\rangle\right).
\end{array}\label{eq:final outcome corresponding relation}
\end{equation}
The first row of the map described above implies that if Alice's (Bob's)
Bell measurement yields $|\phi^{+}\rangle$ then Alice (Bob) stores
00 without announcement for future use in implementing the protocol.
Other rows of the above equation follow similar mapping.

Now, we briefly describe the idea in the background of our CQKA protocol.
Initially, Alice and Bob inform Charlie to start the protocol. Charlie prepares a sequence in one of the two states, $|\phi^{+}\rangle_{{\rm C_{1}C_{2}}}$
and $|\phi^{-}\rangle_{{\rm C_{1}C_{2}}}$. For instance, we consider the state $|\phi^{+}\rangle_{{\rm C_{1}C_{2}}}$. This preparation is performed after receiving a request from Alice and Bob. Here, the subscripts  ${\rm C_{1}}$ and ${\rm C_{2}}$ denote the first and second qubits of Charlie's Bell state, respectively. Charlie notes the classical sequence $k_{C}$ (here $0$) which stores the classical information of his prepared Bell states. Then, he sends
qubit ${\rm C_{1}}$ (${\rm C_{2}}$) to Alice (Bob) and keeps $k_{C}$
secret. After getting the qubit ${\rm C_{1}}$ (${\rm C_{2}}$) from
Charlie, Alice (Bob) prepares her (his) own qubit in the computational
basis ( i.e., $Z\in\left\{ |0\rangle,|1\rangle\right\} $) say, $|0\rangle$
($|1\rangle$). Alice notes the classical sequence $k_{A}$ (here
$0$) and keeps it secret. We can recognize Alice's (Bob's) prepared
qubit with subscript ${\rm A}$ (${\rm B}$) or qubit ${\rm A}$ (${\rm B}$).
Alice (Bob) performs the CNOT operation on target qubit ${\rm A}$
(${\rm B}$) with the control qubit ${\rm C_{1}}$ (${\rm C_{2}}$).
The following equation represents the scenario described above

\begin{equation}
\begin{array}{lcl}
|\psi_{2}\rangle & = & \frac{1}{\sqrt{2}}{\rm CNOT_{C_{1}\rightarrow A}CNOT_{C_{2}\rightarrow B}}\left(|0\rangle_{{\rm A}}|\phi^{+}\rangle_{{\rm C_{1}C_{2}}}|1\rangle_{{\rm B}}\right)\\
 & = & \frac{1}{\sqrt{2}}\left(|00\rangle_{{\rm AC_{1}}}|01\rangle_{{\rm C_{2}B}}+|11\rangle_{{\rm AC_{1}}}|10\rangle_{{\rm C_{2}B}}\right).
\end{array}\label{eq:after=000020CNOT=000020operation}
\end{equation}
In the present situation, Alice and Bob, each have two qubits ${\rm C_{1},A}$
and ${\rm C_{2},B}$, respectively. Both parties perform measurement
using the Bell basis. After executing the Bell measurement they have
the resulting state described as follows

\begin{equation}
\begin{array}{lcl}
|\psi_{2}\rangle & = & \frac{1}{\sqrt{2}}\left(|\phi^{+}\rangle_{{\rm AC_{1}}}|\psi^{+}\rangle_{{\rm C_{2}B}}+|\phi^{-}\rangle_{{\rm AC_{1}}}|\psi^{-}\rangle_{{\rm C_{2}B}}\right).\end{array}\label{eq:=00005CPsi2}
\end{equation}

After getting the measurement result, they request Charlie to announce
his bit sequence $k_{C}$ (here $0$) that corresponds to his prepared
states (using mapping of Eq. (\ref{eq:Charlie's corresponding relation})).
Also, Alice (Bob) announces her (his) bit sequence $k_{A}$ ($k_{B}$)
depending upon her prepared state (his measurement result on qubit
${\rm C_{2}}$ and qubit ${\rm B}$ after Bell measurement) according to Eq. $(\ref{eq:Alice's corresponding relation})$
$\left(\text{Eq}\text{. } (\ref{eq:Bob's corresponding relation})\right)$.
In our case, Alice announces bit value $0$, and Bob announces bit
value $0$ or $1$ with $\frac{1}{2}$ probability as he could get
the measurement outcome $|\psi^{-}\rangle_{{\rm C_{2}B}}$
or $|\psi^{+}\rangle_{{\rm C_{2}B}}$
with equal probability (see Eq. (\ref{eq:=00005CPsi2})). The mapping
among their (Charlie, Alice, and Bob) announcements, self-measurement
outcome (not disclosed), and other legitimate party's measurement
outcome (not disclosed) are illustrated in the Table \ref{tab:Relation of final key and announcement for P1}.
Let us take one combination; bit values announced by Charlie, Alice,
and Bob are $0,0$, and $1$; As Alice and Bob already know two-bit
classical information of their own measurement results (by Eq $.(\ref{eq:final outcome corresponding relation})$),
Alice (Bob) guesses Bob's (Alice's) measurement outcome is $10$ ($00$)
according to the Table \ref{tab:Relation of final key and announcement for P1}.
To get the final agreement key after executing the protocol, Alice
and Bob do modulo 2 addition between corresponding bit values $\left(\text{by Eq. } (\ref{eq:final outcome corresponding relation})\right)$
of their measurement outcomes and get a two-bit result; Again do modulo
2 addition between the first and second bit of this two-bit result.
In our example, it is $10\oplus00=10$, then final key is $1\oplus0=1$.

It should be noted that there are a total of eight different combinations
of states due to individual choices by Alice, Bob and Charlie. These
different states may be denoted by $|\psi_{i}\rangle,i\in\left\{ 1,2,\cdots,8\right\} $.
In our case $|\psi_{2}\rangle$ is already discussed, the states of
the rest of the combination are expressed as follows,

\begin{equation}
\begin{array}{lcl}
|\psi_{1}\rangle & = & \frac{1}{\sqrt{2}}{\rm CNOT}{}_{{\rm C_{1}\rightarrow{\rm A}}}{\rm CNOT_{C_{2}\rightarrow B}}\left(|0\rangle_{{\rm A}}|\phi^{+}\rangle_{{\rm C_{1}C_{2}}}|0\rangle_{{\rm B}}\right)\\
 & = & \frac{1}{\sqrt{2}}\left(|\phi^{+}\rangle_{{\rm AC_{1}}}|\phi^{+}\rangle_{{\rm C_{2}B}}+|\phi^{-}\rangle_{{\rm AC_{1}}}|\phi^{-}\rangle_{{\rm C_{2}B}}\right),
\end{array}\label{eq:=00005CPsi1}
\end{equation}

\begin{equation}
\begin{array}{lcl}
|\psi_{3}\rangle & = & \frac{1}{\sqrt{2}}{\rm CNOT}{}_{{\rm C_{1}\rightarrow{\rm A}}}{\rm CNOT_{C_{2}\rightarrow B}}\left(|1\rangle_{{\rm A}}|\phi^{+}\rangle_{{\rm C_{1}C_{2}}}|0\rangle_{{\rm B}}\right)\\
 & = & \frac{1}{\sqrt{2}}\left(|\psi^{+}\rangle_{{\rm AC_{1}}}|\phi^{+}\rangle_{{\rm C_{2}B}}-|\psi^{-}\rangle_{{\rm AC_{1}}}|\phi^{-}\rangle_{{\rm C_{2}B}}\right),
\end{array}\label{eq:=00005CPsi3}
\end{equation}

\begin{equation}
\begin{array}{lcl}
|\psi_{4}\rangle & = & \frac{1}{\sqrt{2}}{\rm CNOT}{}_{{\rm C_{1}\rightarrow{\rm A}}}{\rm CNOT_{C_{2}\rightarrow B}}\left(|1\rangle_{{\rm A}}|\phi^{+}\rangle_{{\rm C_{1}C_{2}}}|1\rangle_{{\rm B}}\right)\\
 & = & \frac{1}{\sqrt{2}}\left(|\psi^{+}\rangle_{{\rm AC_{1}}}|\psi^{+}\rangle_{{\rm C_{2}B}}-|\psi^{-}\rangle_{{\rm AC_{1}}}|\psi^{-}\rangle_{{\rm C_{2}B}}\right),
\end{array}\label{eq:=00005CPsi4}
\end{equation}

\begin{equation}
\begin{array}{lcl}
|\psi_{5}\rangle & = & {\rm \frac{1}{\sqrt{2}}CNOT{}_{C_{1}\rightarrow{\rm A}}CNOT_{C_{2}\rightarrow B}}\left(|0\rangle_{{\rm A}}|\phi^{-}\rangle_{{\rm C_{1}C_{2}}}|0\rangle_{{\rm B}}\right)\\
 & = & \frac{1}{\sqrt{2}}\left(|\phi^{+}\rangle_{{\rm AC_{1}}}|\phi^{-}\rangle_{{\rm C_{2}B}}+|\phi^{-}\rangle_{{\rm AC_{1}}}|\phi^{+}\rangle_{{\rm C_{2}B}}\right),
\end{array}\label{eq:=00005CPsi5}
\end{equation}

\begin{equation}
\begin{array}{lcl}
|\psi_{6}\rangle & = & \frac{1}{\sqrt{2}}{\rm CNOT_{C_{1}\rightarrow{\rm A}}CNOT_{C_{2}\rightarrow B}}\left(|0\rangle_{{\rm A}}|\phi^{-}\rangle_{{\rm C_{1}C_{2}}}|1\rangle_{{\rm B}}\right)\\
 & = & \frac{1}{\sqrt{2}}\left(|\phi^{+}\rangle_{{\rm AC_{1}}}|\psi^{-}\rangle_{{\rm C_{2}B}}+|\phi^{-}\rangle_{{\rm AC_{1}}}|\psi^{+}\rangle_{{\rm C_{2}B}}\right),
\end{array}\label{eq:=00005CPsi6}
\end{equation}

\begin{equation}
\begin{array}{lcl}
|\psi_{7}\rangle & = & \frac{1}{\sqrt{2}}{\rm CNOT_{C_{1}\rightarrow{\rm A}}CNOT_{C_{2}\rightarrow B}}\left(|1\rangle_{{\rm A}}|\phi^{-}\rangle_{{\rm C_{1}C_{2}}}|0\rangle_{{\rm B}}\right)\\
 & = & \frac{1}{\sqrt{2}}\left(|\psi^{+}\rangle_{{\rm AC_{1}}}|\phi^{-}\rangle_{{\rm C_{2}B}}-|\psi^{-}\rangle_{{\rm AC_{1}}}|\phi^{+}\rangle_{{\rm C_{2}B}}\right),
\end{array}\label{eq:=00005CPsi7}
\end{equation}

\begin{equation}
\begin{array}{lcl}
|\psi_{8}\rangle & = & \frac{1}{\sqrt{2}}{\rm CNOT_{C_{1}\rightarrow{\rm A}}CNOT_{C_{2}\rightarrow B}}\left(|1\rangle_{{\rm A}}|\phi^{-}\rangle_{{\rm C_{1}C_{2}}}|1\rangle_{{\rm B}}\right)\\
 & = & \frac{1}{\sqrt{2}}\left(|\psi^{+}\rangle_{{\rm AC_{1}}}|\psi^{-}\rangle_{{\rm C_{2}B}}-|\psi^{-}\rangle_{{\rm AC_{1}}}|\psi^{+}\rangle_{{\rm C_{2}B}}\right).
\end{array}\label{eq:=00005CPsi8}
\end{equation}

\begin{center}
\begin{table}
\caption{\label{tab:Relation of final key and announcement for P1}This table
shows the relation between the announcement by Charlie, Alice and
Bob with the final key.}

\centering{}%
\begin{tabular*}{16.5cm}{@{\extracolsep{\fill}}|>{\raggedright}p{2cm}|>{\raggedright}p{2cm}|>{\raggedright}p{2cm}|>{\raggedright}p{2.5cm}|>{\raggedright}p{2.5cm}|>{\raggedright}p{2cm}|>{\raggedright}p{2cm}|}
\hline 
\centering{}Announced bit by Charlie $(k_{C})$ & \centering{}Announced bit by Alice $(k_{A})$ & \centering{}Announced bit by Bob $(k_{B})$ & \centering{}Alice's measurement out come guessed by Bob $(r_{A})$ & \centering{}Bob's measurement out come guessed by Alice $(r_{B})$ & \begin{centering}
Value of
\par\end{centering}
\centering{}$r_{A}\oplus r_{B}$ & \centering{}Final key after QKA $(K)$\tabularnewline
\hline 
\centering{}0 & \centering{}0 & \centering{}0 & \centering{}00 & \centering{}00 & \centering{}00 & \centering{}0\tabularnewline
\centering{} & \centering{}0 & \centering{}1 & \centering{}01 & \centering{}01 & \centering{}00 & \centering{}0\tabularnewline
\centering{} & \centering{}0 & \centering{}1 & \centering{}00 & \centering{}10 & \centering{}10 & \centering{}1\tabularnewline
\centering{} & \centering{}0 & \centering{}0 & \centering{}01 & \centering{}11 & \centering{}10 & \centering{}1\tabularnewline
\cline{2-7} \cline{3-7} \cline{4-7} \cline{5-7} \cline{6-7} \cline{7-7} 
\centering{} & \centering{}1 & \centering{}0 & \centering{}10 & \centering{}00 & \centering{}10 & \centering{}1\tabularnewline
\centering{} & \centering{}1 & \centering{}1 & \centering{}11 & \centering{}01 & \centering{}10 & \centering{}1\tabularnewline
\centering{} & \centering{}1 & \centering{}1 & \centering{}10 & \centering{}10 & \centering{}00 & \centering{}0\tabularnewline
\centering{} & \centering{}1 & \centering{}0 & \centering{}11 & \centering{}11 & \centering{}00 & \centering{}0\tabularnewline
\hline 
\centering{}1 & \centering{}0 & \centering{}1 & \centering{}00 & \centering{}01 & \centering{}01 & \centering{}1\tabularnewline
\centering{} & \centering{}0 & \centering{}0 & \centering{}01 & \centering{}00 & \centering{}01 & \centering{}1\tabularnewline
\centering{} & \centering{}0 & \centering{}0 & \centering{}00 & \centering{}11 & \centering{}11 & \centering{}0\tabularnewline
\centering{} & \centering{}0 & \centering{}1 & \centering{}01 & \centering{}10 & \centering{}11 & \centering{}0\tabularnewline
\cline{2-7} \cline{3-7} \cline{4-7} \cline{5-7} \cline{6-7} \cline{7-7} 
\centering{} & \centering{}1 & \centering{}1 & \centering{}10 & \centering{}01 & \centering{}11 & \centering{}0\tabularnewline
\centering{} & \centering{}1 & \centering{}0 & \centering{}11 & \centering{}00 & \centering{}11 & \centering{}0\tabularnewline
\centering{} & \centering{}1 & \centering{}0 & \centering{}10 & \centering{}11 & \centering{}01 & \centering{}1\tabularnewline
\centering{} & \centering{}1 & \centering{}1 & \centering{}11 & \centering{}10 & \centering{}01 & \centering{}1\tabularnewline
\hline 
\end{tabular*}
\end{table}
\par\end{center}

\subsection*{Proposed Protocol 1 for CQKA}

\begin{figure}
\begin{centering}
\includegraphics[scale=0.4]{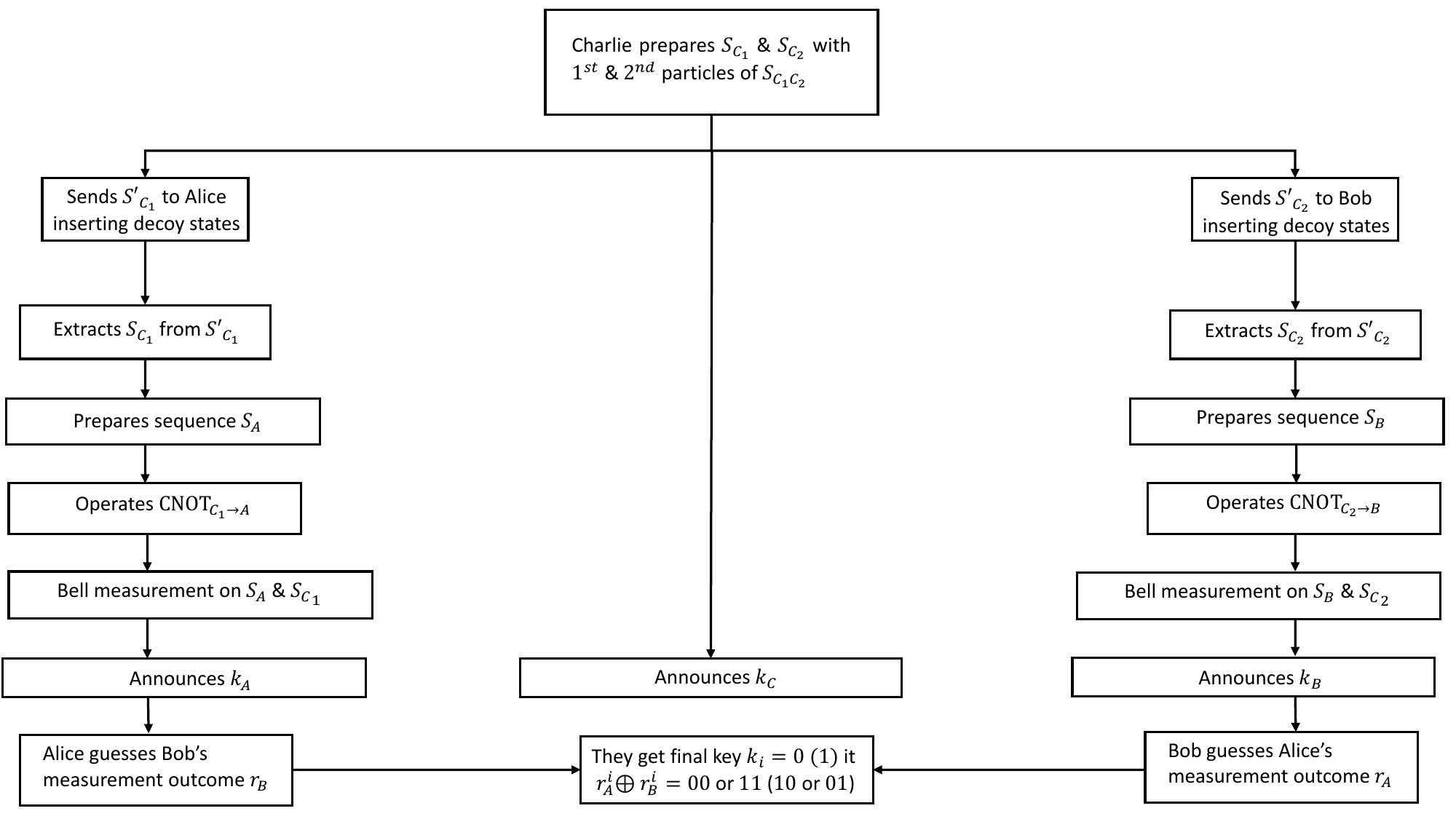}
\par\end{centering}
\caption{\label{fig:This-flowchart-depicting}This flowchart is depicting the
mechanism of the proposed CQKA Protocol 1.}	
\end{figure}

Here, we illustrate the new CQKA protocol in a step-wise manner. 
\begin{description}
\item [{Step~1}] Charlie prepares a sequence of
$n$ Bell states $\left\{ S_{C_{1}C_{2}}\right\} $, and divides it
into two sequences $S_{C_{1}}$ and $S_{C_{2}}$ with first and second
qubits of the Bell states, respectively, leading to
\[
\begin{array}{c}
S_{C_{1}}=\left\{ \vert s\rangle_{C_{1}}^{1},|s\rangle_{C_{1}}^{2},|s\rangle_{C_{1}}^{3},\cdots,|s\rangle_{C_{1}}^{i},\cdots,|s\rangle_{C_{1}}^{n}\right\} ,\\
S_{C_{2}}=\left\{ \vert s\rangle_{C_{2}}^{1},|s\rangle_{C_{2}}^{2},|s\rangle_{C_{2}}^{3},\cdots,|s\rangle_{C_{2}}^{i},\cdots,|s\rangle_{C_{2}}^{n}\right\} .
\end{array}
\]
Here, the subscripts ${\rm C_{1}}$ and ${\rm C_{2}}$ represent the
first and second qubits, respectively, of Charlie\textquoteright s
Bell state sequence $S_{C_{1}C_{2}}$. Each Bell state in the sequence
is either $|\phi^{+}\rangle$ or $|\phi^{-}\rangle$, and the superscript
(e.g., $\left\{ 1,2,3,\cdots,i,\cdots,n\right\} $) indicates the
position of the qubits in $S_{C_{1}}$ and $S_{C_{2}}$.
\item [{Step~2}] Charlie then randomly
prepares $2p$ decoy qubits using the computational basis ($Z$ basis,
i.e., $\left\{ |0\rangle,|1\rangle\right\} $\}) or the diagonal basis
($X$ basis, i.e., $\left\{ |+\rangle,|-\rangle\right\} $). He inserts
$p$ randomly prepared decoy qubits in arbitrary positions in $S_{C_{1}}$
and another $p$ decoy qubits in arbitrary positions in $S_{C_{2}}$.
This results in enlarged sequences $S_{C_{1}}^{\prime}$ and $S_{C_{2}}^{\prime}$.
Charlie sends the enlarged sequences $S_{C_{1}}^{\prime}$ and $S_{C_{2}}^{\prime}$
to Alice and Bob, respectively, while retaining a classical bit sequence
$k_{C}$. This sequence allows Charlie to prepare the original Bell
states in $S_{C_{1}C_{2}}$. \\
The relationship between $k_{C}$ and $S_{C_{1}C_{2}}$
follows a defined correlation described by a governing Eq$.$ (\ref{eq:Charlie's corresponding relation}).
\item [{Step~3}] Alice and Bob each
receive sequences $S_{C_{1}}^{\prime}$ and $S_{C_{2}}^{\prime}$
from Charlie, respectively. They check the decoy qubits for security
and obtain $S_{C_{1}}$ and $S_{C_{2}}$. \\
Charlie publicly announces the positions and encoding
used for the decoy states. Alice and Bob then verify the channel's
security by computing the error rate. If the error rate is found below
the tolerable error limit, they proceed.
\item [{Step~4}] Alice and Bob prepare
their single-qubit sequences (of length $n$), $S_{A}$ and $S_{B}$,
using the $Z$-basis. Alice records the classical bit sequence $k_{A}$
used to generate $S_{A}$ based on a defined relation (Eq. (\ref{eq:Alice's corresponding relation})).
Here, $S_{A}$ and $S_{B}$ can be visualized as the sequences
\[
\begin{array}{lcl}
S_{A} & = & \left\{ |s\rangle_{A}^{1},|s\rangle_{A}^{2},|s\rangle_{A}^{3},\cdots,|s\rangle_{A}^{i},\cdots,|s\rangle_{A}^{n}\right\} ,\\
S_{B} & = & \left\{ |s\rangle_{B}^{1},|s\rangle_{B}^{2},|s\rangle_{B}^{3},\cdots,|s\rangle_{B}^{i},\cdots,|s\rangle_{B}^{n}\right\} .
\end{array}
\]

\item [{Step~5}] Both Alice and Bob
perform a CNOT operation, using each qubit in $S_{A}$ and $S_{B}$
as the target qubit and the corresponding qubit in $S_{C_{1}}$ and
$S_{C_{2}}$ as the control qubit, respectively. This results in a
correlated two-qubit system with pairs $({\rm C_{1},A)}$ for Alice
and ${\rm (C_{2},B})$ for Bob.
\item [{Step~6}] Alice and Bob each
perform a Bell-state measurement on their two-qubit systems and obtain
results $r_{A}$ and $r_{B},$ respectively (using Eq. (\ref{eq:final outcome corresponding relation})).
\\
They keep these results private. Bob also records
a classical bit sequence $k_{B}$ derived from his measurement results
using the correlation defined in Eq. (\ref{eq:Bob's corresponding relation}).
\item [{Step~7}] Alice and Bob request
Charlie to publicly announce his classical bit sequence $k_{C}$. 
\item [{Step~8}] After Charlie's announcement,
Alice and Bob also share their classical bit sequences $k_{A}$ and
$k_{B}$ publicly.
\item [{Step~9}] Using the mapping
provided in Table \ref{tab:Relation of final key and announcement for P1},
Alice predicts Bob's bit sequence $r_{B}$, and Bob predicts Alice's
bit sequence $r_{A}$. They then perform modulo 2 addition between
corresponding elements of $r_{A}$ and $r_{B}$. If the result (by
modulo $2$ addition) is $00$ or $11$, the final key bit $K_{i}=0$;
if it is $01$ or $10$, then $K_{i}=1$.
\end{description}
Alice and Bob, both take a subset of the final key $\left(K_{i}\right)$
and compare it with each other. If they get key mismatch probability
or error rate is more than a tolerable error limit, they abort the
protocol or else accept the final key for their secure communication
purpose after post-processing steps involving error correction and
privacy amplification. The CQKA protocol is also presented through
a flowchart shown in Figure \ref{fig:This-flowchart-depicting}.

Here, we will explore a different approach to our CQKA protocol by
eliminating the role of the controller and our scheme will reduce
to a two-party QKA protocol. This approach leads to the scheme being
more feasible and less noisy in comparison to the previous one. We
will elaborately discuss these two schemes in the context of merit
and demerit with some previously proposed schemes in Section \ref{sec:Comparison}.
One can use one of these two schemes in keeping with mind on demands
of various situations. This simple modification for the two-party
QKA protocol is described in the following steps without including
the decoy state to simplify our description and the mapping of Alice
and Bob's announcement and the final key is shown in Table \ref{tab:Relation of final key and announcement for P2}.

\subsection*{Proposed Protocol 2 for QKA}
\begin{description}
\item [{Step~1}] To begin with, Alice prepares
$n$ Bell states $\left\{ S_{C_{1}C_{2}}\right\} $, and splits them
into two sequences: $S_{C_{1}}$ (a sequence with all the first qubits
of the Bell states) and $S_{C_{2}}$ (a sequence with all the second
qubits of the Bell states). She also prepares a sequence of $n$ single
qubits ($S_{A})$ in the $Z$-basis, ensuring that the bit values
correspond to $k_{C}$ and $k_{A}$ will be the same\footnote{Following the relations defined in Eqs. (\ref{eq:Charlie's corresponding relation},\ref{eq:Alice's corresponding relation}
and \ref{eq:Bob's corresponding relation})). These relations
apply to Protocol 2 as well.}. The sequences prepared by Alice are,
\[
\begin{array}{c}
S_{C_{1}}=\left\{ \vert s\rangle_{C_{1}}^{1},|s\rangle_{C_{1}}^{2},|s\rangle_{C_{1}}^{3},\cdots,|s\rangle_{C_{1}}^{i},\cdots,|s\rangle_{C_{1}}^{n}\right\} ,\\
S_{C_{2}}=\left\{ \vert s\rangle_{C_{2}}^{1},|s\rangle_{C_{2}}^{2},|s\rangle_{C_{2}}^{3},\cdots,|s\rangle_{C_{2}}^{i},\cdots,|s\rangle_{C_{2}}^{n}\right\} ,\\
S_{A}=\left\{ |s\rangle_{A}^{1},|s\rangle_{A}^{2},|s\rangle_{A}^{3},\cdots,|s\rangle_{A}^{i},\cdots,|s\rangle_{A}^{n}\right\} .
\end{array}
\]

\item [{Step~2}] Alice keeps the $S_{C_{1}}$
sequence and sends $S_{C_{2}}$ sequence to Bob. 
\item [{Step~3}] Meanwhile, Bob independently
prepares a sequence of $n$ single qubits $S_{B}$ in the $Z$-basis
as
\[
S_{B}=\left\{ |s\rangle_{B}^{1},|s\rangle_{B}^{2},|s\rangle_{B}^{3},\cdots,|s\rangle_{B}^{i},\cdots,|s\rangle_{B}^{n}\right\} .
\]

\item [{Step~4}] Same as Step 5 of Protocol 1.
\item [{Step~5}] Same as Step 6 of Protocol 1.
\item [{Step~6}] Since $k_{A}$ and $k_{C}$ are identical,
Alice and Bob announce their sequences $k_{A}$ and $k_{B}$ publicly.
\item [{Step~7}] Same as Step 9 of Protocol 1.
\end{description}
\begin{center}
\begin{table}
\caption{\label{tab:Relation of final key and announcement for P2}This shows
the relation between the announcement by Alice, and Bob with their
final key.}

\centering{}%
\begin{tabular*}{16.5cm}{@{\extracolsep{\fill}}|>{\raggedright}p{2.5cm}|>{\raggedright}p{2.5cm}|>{\raggedright}p{2.5cm}|>{\raggedright}p{2.5cm}|>{\raggedright}p{2.5cm}|>{\raggedright}p{2.5cm}|}
\hline 
\centering{}Announced bit by Alice $(k_{A})$ & \centering{}Announced bit by Bob $(k_{B})$ & \centering{}Alice's measurement out come guessed by Bob $(r_{A})$ & \centering{}Bob's measurement out come guessed by Alice $(r_{B})$ & \begin{centering}
Value of
\par\end{centering}
\centering{}$r_{A}\oplus r_{B}$ & \centering{}Final key after QKA $(K)$\tabularnewline
\hline 
\centering{}0 & \centering{}0 & \centering{}00 & \centering{}00 & \centering{}00 & \centering{}0\tabularnewline
\centering{}0 & \centering{}1 & \centering{}01 & \centering{}01 & \centering{}00 & \centering{}0\tabularnewline
\centering{}0 & \centering{}1 & \centering{}00 & \centering{}10 & \centering{}10 & \centering{}1\tabularnewline
\centering{}0 & \centering{}0 & \centering{}01 & \centering{}11 & \centering{}10 & \centering{}1\tabularnewline
\hline 
\centering{}1 & \centering{}1 & \centering{}10 & \centering{}01 & \centering{}11 & \centering{}0\tabularnewline
\centering{}1 & \centering{}0 & \centering{}11 & \centering{}00 & \centering{}11 & \centering{}0\tabularnewline
\centering{}1 & \centering{}0 & \centering{}10 & \centering{}11 & \centering{}01 & \centering{}1\tabularnewline
\centering{}1 & \centering{}1 & \centering{}11 & \centering{}10 & \centering{}01 & \centering{}1\tabularnewline
\hline 
\end{tabular*}
\end{table}
\par\end{center}

\section{Security analysis of the proposed CQKA protocols\label{sec:Security-Analysis}}

With the other important conditions, \emph{security} also acts a very
important role not only in QKA protocol but also in any quantum communication
protocol. In this section, we analyze the security against impersonated
fraudulent attack, and more general collective attack on the proposed
Protocol 1. These two security analyses also be applicable to the
Protocol 2. We also show that our protocols are secure with a small
size length of the final agreement key $({\rm n=6})$. 

\subsection{Security analysis opposed to impersonated fraudulent attack for the
proposed protocol}

To analyze this attack we may consider that Eve tries to impersonate
Charlie by using a generalized two-qubit system. Without loss of generality,
we ignore the security provided by decoy qubits. In Step~2 of our
protocol, Eve who impersonates may transmit the first, and the second
qubits of her generated two-qubit system to Alice and Bob like Charlie.
First, we consider the situation in which Alice and Bob choose the
same state $|0\rangle$ as an element of sequences $S_{A}$
and $S_{B}$ at their end, and they proceed to execute further steps
as the protocol describes. We can write the following equation after
the CNOT operation is performed by both parties,

\begin{equation}
\begin{array}{lcl}
|\Phi_{1}\rangle & = & {\rm CNOT_{C_{1}\rightarrow A}CNOT_{C_{2}\rightarrow B}}\left(|0\rangle_{{\rm A}}\left({\rm {\rm a}|00\rangle+b|01\rangle+c|10\rangle+d|11\rangle}\right)_{{\rm C_{1}C_{2}}}|0\rangle_{{\rm B}}\right)\\
 & = & \left({\rm a|0\rangle|00\rangle|0\rangle+b|0\rangle|01\rangle|1\rangle+c|1\rangle|10\rangle|0\rangle+d|1\rangle|11\rangle|1\rangle}\right)_{{\rm AC_{1}C_{2}B}},
\end{array}\label{eq:Impersonate fraudulent after CNOT}
\end{equation}
where ${\rm a,b,c}$, and ${\rm d}$ are the probability amplitude
of the states $|00\rangle,|01\rangle,|10\rangle$, and $|11\rangle$
of Eve's two-qubit system\footnote{One can consider the well-known two-qubit state prepared by Eve, and
check the robustness against impersonated fraudulent attack by putting
the desired value of these non-zero constants ${\rm a,b,c}$, and
${\rm d}$.}; with the normalization condition ${\rm |a|^{2}+|b|^{2}+|c|^{2}+|d|^{2}=1}$.
 Now, Alice
and Bob measure their two-qubit system ${\rm A,C_{1}}$ and ${\rm C_{2},B}$,
respectively in the Bell basis. For the convenience of visualizing
the result after Bell measurements, we may express the above equation
in Bell basis as

\begin{equation}
\begin{array}{lcl}
|\Phi_{1}\rangle & = & \frac{1}{2}\left[{\rm a}\left(|\phi^{+}\rangle|\phi^{+}\rangle+|\phi^{+}\rangle|\phi^{-}\rangle+|\phi^{-}\rangle|\phi^{+}\rangle+|\phi^{-}\rangle|\phi^{-}\rangle\right)\right.\\
 & + & {\rm b\left(|\phi^{+}\rangle|\phi^{+}\rangle-|\phi^{+}\rangle|\phi^{-}\rangle+|\phi^{-}\rangle|\phi^{+}\rangle-|\phi^{-}\rangle|\phi^{-}\rangle\right)}\\
 & + & {\rm c}\left(|\phi^{+}\rangle|\phi^{+}\rangle+|\phi^{+}\rangle|\phi^{-}\rangle-|\phi^{-}\rangle|\phi^{+}\rangle-|\phi^{-}\rangle|\phi^{-}\rangle\right)\\
 & + & {\rm \left.d\left(|\phi^{+}\rangle|\phi^{+}\rangle-|\phi^{+}\rangle|\phi^{-}\rangle-|\phi^{-}\rangle|\phi^{+}\rangle+|\phi^{-}\rangle|\phi^{-}\rangle\right)\right]_{AC_{1}C_{2}B}}
\end{array}.\label{eq:Impersonate fraudulent Bell state form 1}
\end{equation}

Here, Eq$.$ (\ref{eq:Impersonate fraudulent Bell state form 1})
shows the composite system of Alice, Bob, and Eve after impersonate
attack of Eve through the generation of a two-qubit system. We can
calculate the detection probability of Eve by taking the comparison
with Eq$.$ (\ref{eq:=00005CPsi1}) which represents the expected outcome
by Alice and Bob. After necessary evaluation, the probability of detecting
Eve is obtained as follows, 

\begin{equation}
\begin{array}{lcl}
{\rm P_{|0\rangle|\phi^{+}\rangle|0\rangle}} & = & \frac{1}{2}\left[{\rm \left(|a|^{2}+|b|^{2}+|c|^{2}+|d|^{2}\right)-\left(a^{*}b+b^{*}c+c^{*}b+d^{*}a\right)}\right]\\
 & = & \frac{1}{2}\left[{\rm 1-\left(a^{*}b+b^{*}c+c^{*}b+d^{*}a\right)}\right],
\end{array}\label{eq:first detection probability}
\end{equation}
and similarly,

\begin{equation}
\begin{array}{lcl}
{\rm P_{|0\rangle|\phi^{-}\rangle|0\rangle}} & = & \frac{1}{2}\left[{\rm \left(|a|^{2}+|b|^{2}+|c|^{2}+|d|^{2}\right)+\left(a^{*}b+b^{*}c+c^{*}b+d^{*}a\right)}\right]\\
 & = & \frac{1}{2}\left[{\rm 1+\left(a^{*}b+b^{*}c+c^{*}b+d^{*}a\right)}\right].
\end{array}\label{eq:fifth detection probability}
\end{equation}
Here, subscripts $|0\rangle,|\phi^{+}\rangle$, and $|0\rangle$ represent
states which are prepared by Alice, Bob, and Charlie, respectively.
If we calculate the other situations, i.e., Alice and Bob prepare
states with other combinations  (e.g., $|0\rangle_{{\rm A}},|1\rangle_{{\rm B}};|1\rangle_{{\rm A}},|0\rangle_{{\rm B}}$,
and $|1\rangle_{{\rm A}},|1\rangle_{{\rm B}}$), the detection probabilities will
be,

\begin{equation}
\begin{array}{lcl}
P_{|0\rangle|\phi^{-}\rangle|0\rangle}=P_{|0\rangle|\phi^{+}\rangle|1\rangle} & = & P_{|1\rangle|\phi^{+}\rangle|0\rangle}={\rm P_{|1\rangle|\phi^{+}\rangle|1\rangle},}\\
{\rm P_{|0\rangle|\phi^{-}\rangle|0\rangle}=P_{|0\rangle|\phi^{-}\rangle|1\rangle}} & = & P_{|1\rangle|\phi^{-}\rangle|0\rangle}={\rm P_{|1\rangle|\phi^{-}\rangle|1\rangle}.}
\end{array}\label{eq:all detection probabilities}
\end{equation}
Consequently, the detection possibility for each transmission reads
as,

\begin{equation}
\begin{array}{lcl}
{\rm P_{d}} & = & \frac{1}{8}\left[{\rm P_{|0\rangle|\phi^{+}\rangle|0\rangle}}+{\rm P_{|0\rangle|\phi^{+}\rangle|1\rangle}+{\rm P_{|1\rangle|\phi^{+}\rangle|0\rangle}+{\rm P_{|1\rangle|\phi^{+}\rangle|1\rangle}}}}\right.\\
 & + & \left.{\rm P_{|0\rangle|\phi^{-}\rangle|0\rangle}}+{\rm P_{|0\rangle|\phi^{-}\rangle|1\rangle}+{\rm P_{|1\rangle|\phi^{-}\rangle|0\rangle}+{\rm P_{|1\rangle|\phi^{-}\rangle|1\rangle}}}}\right]\\
 & = & \frac{1}{2}\left({\rm |a|^{2}+|b|^{2}+|c|^{2}+|d|^{2}}\right)\\
 & = & \frac{1}{2}.
\end{array}\label{eq:final detection probability of EVE}
\end{equation}
 A detection probability of $\frac{1}{2}$ represents the maximum statistical randomness
achievable for detecting the presence of an adversary, thereby ensuring
the protocol's security. Mathematically, this implies that for each
message transmission, there is a $0.5$ probability of detecting an
attacker (Eve) attempting to tamper with or forge a message. According
to Simmons theorem \citep{S_88}, a protocol is secure if the adversary
cannot reduce the detection probability below the randomness threshold
of $\frac{1}{2}$ without being detected. This is because $\frac{1}{2}$
corresponds to the maximum entropy for a binary detection event (detected
or undetected), ensuring that Eve cannot deterministically predict
or evade detection. If Eve makes $n$ independent tampering attempts
(in $n$ transmissions), the probability of escaping detection is
$\left(\frac{1}{2}\right)^{n}$, which decreases exponentially as
$n$ increases, virtually guaranteeing detection with repeated attempts.
Consequently, the inherent randomness associated with a detection
probability of $\frac{1}{2}$ ensures unpredictability for the adversary
and meets Simmons security criteria, providing a robust defense against
tampering and forgery. Using Simmons theory \citep{S_88}, the proposed
protocol is unconditionally secure against impersonated fraudulent
attack using Eq$.$ (\ref{eq:final detection probability of EVE}).

\subsection{Security analysis opposed to collective attack strategy
on two-way channel}

There are three classes of attacks which an eavesdropper
can perform on the schemes for secure quantum communication. These
classes are referred to as: (i) Individual attacks, (ii) Collective
attacks, and (iii) Coherent attacks. Among these class of attacks,
individual attacks are the weakest. In individual attacks, Eve interacts
with each of Alice's signal systems separately. For each signal system,
Eve attaches an auxiliary system and applies a fixed unitary operation.
Subsequently, Eve measures each system individually right after the
sifting step (i.e., before Alice and Bob perform classical processing).
Collective attacks \citep{BM97,BM+97} are similar, except the fact
that Eve may delay her measurement until the end of the protocol.
Her measurement choice may depend on the messages exchanged by Alice
and Bob for error correction and privacy amplification, and she may
measure all auxiliary systems jointly. Here, Eve is more powerful
than an eavesdropper who can only perform individual attack as she
is capable of performing joint measurement. Analyzing a protocol's
security against collective attacks is a step towards proving security
against coherent attacks, the most general case. In this section,
we analyze security against collective attacks to determine the tolerable
error limit under the security threshold. We first describe Eve's
attack strategy using a two-way quantum channel with her non-orthogonal
ancilla states and calculate the detection probability of Eve's presence.
We also analyze the probability of Eve obtaining information, considering
parameters such as Eve's detection probability and the final key information.
Furthermore, we use the asymptotic secret key rate, as defined by
Devetak and Winter \citep{DW05}, for collective attacks to get the
secure bound \citep{RGK_05}.

In our proposed CQKA protocol, there exist two quantum channels, i.e.,
Charlie-Alice (CA) channel and Charlie-Bob (CB) channel. Assume these
two channels are available for Eve to attack to get final key information
with minimum detection possibility. 
In this type of attack, Eve prepares two \emph{ancillae,}
denoted as $|\zeta\rangle$ and $|\eta\rangle$. These ancillae constitute
a two-qubit system intended for attacking two specific channels. At first, Eve intercepts the qubit (qubit
${\rm C_{1}}$) from the CA channel and performs an entangling operation
${\rm E_{C_{1}}}$ on qubit ${\rm C_{1}}$ which would entangle qubit
${\rm C_{1}}$ and the \emph{ancilla} $|\zeta\rangle$. After this operation, Eve sends the qubit to Alice.
In a similar way, Eve performs another entangling operation ${\rm E_{C_{2}}}$
on the qubit (qubit ${\rm C_{2}}$) which would transmit the CB channel,
and that operation would entangle qubit ${\rm C_{2}}$ and her \emph{ancilla
$|\mathcal{\eta}\rangle$. }Eve tries to achieve
the final key by employing the obtained result from measurement of
$|\zeta\rangle$ and $|\eta\rangle$. Here, Eve's operation ${\rm E_{C_{1}}}$
on the qubit that in the CA channel can be represented in a general
expression,

\begin{equation}
{\rm \begin{array}{lcl}
|\rm0_{C_{1}}\zeta\rangle & \stackrel{{\rm E_{C_{1}}}}{\longrightarrow} & A_{\zeta}|\rm0_{C_{1}}\zeta_{00}\rangle+B_{\zeta}|\rm1_{C_{1}}\zeta_{01}\rangle,\\
|\rm0_{C_{1}}\zeta\rangle & \stackrel{{\rm E_{C_{1}}}}{\longrightarrow} & B_{\zeta}|\rm0_{C_{1}}\zeta_{10}\rangle+A_{\zeta}|1_{C_{1}}\zeta_{11}\rangle.
\end{array}}\label{eq:After E1 Operation}
\end{equation}
Here, subscript ${\rm C_{1}}$ (${\rm C_{2}}$) represents a qubit
of channel CA (CB). Similarly, Eve's operation ${\rm E_{C_{2}}}$
on the qubit (qubit ${\rm C_{2}}$) in the CB channel can be expressed
as,

\begin{equation}
{\rm \begin{array}{lcl}
|\rm0_{C_{2}}\eta\rangle & \stackrel{{\rm E_{C_{2}}}}{\longrightarrow} & A_{\eta}|\rm0_{C_{2}}\eta_{00}\rangle+B_{\zeta}|\rm1_{C_{2}}\eta_{01}\rangle,\\
|\rm0_{C_{2}}\eta\rangle & \stackrel{{\rm E_{C_{2}}}}{\longrightarrow} & B_{\eta}|\rm0_{C_{2}}\eta_{10}\rangle+A_{\zeta}|\rm1_{C_{2}}\eta_{11}\rangle.
\end{array}}\label{eq:After E2 Operation}
\end{equation}
Unitary of operators ${\rm E_{C_{1}}}$ and ${\rm E_{C_{2}}}$ demand that
${\rm A_{\zeta}^{2}+B_{\zeta}^{2}=1}$, ${\rm A_{\eta}^{2}+B_{\eta}^{2}=1}$,
$\langle\zeta_{00}|\zeta_{10}\rangle+\langle\zeta_{01}|\zeta_{11}\rangle=0$
and $\langle\eta_{00}|\eta_{10}\rangle+\langle\eta_{01}|\eta_{11}\rangle=0$.
For simplification of our calculations, we may consider a set of assumptions,
$\langle\zeta_{00}|\zeta_{01}\rangle=\langle\zeta_{00}|\zeta_{10}\rangle=\langle\zeta_{10}|\zeta_{11}\rangle=\langle\zeta_{01}|\zeta_{11}\rangle=0$,
and $\langle\eta_{00}|\eta_{01}\rangle=\langle\eta_{00}|\eta_{10}\rangle=\langle\eta_{10}|\eta_{11}\rangle=\langle\eta_{01}|\eta_{11}\rangle=0$.
These assumptions eliminate the generality of Eve's attack, but it
still most typically states after Eve's attacking operation \citep{ZZZX_2006}.
To express Eve's non-orthogonal states we can write $\langle\zeta_{00}|\zeta_{11}\rangle=\cos\alpha_{\zeta},\langle\zeta_{01}|\zeta_{10}\rangle=\cos\beta_{\zeta},\langle\eta_{00}|\eta_{11}\rangle=\cos\alpha_{\eta},\langle\eta_{01}|\eta_{10}\rangle=\cos\beta_{\eta}$. 

Applying the above relations in the condition that represents by Eq$.$ (\ref{eq:=00005CPsi2}),
we can get the non-detection possibility of Eve. First, we obtain
the composite system after Eve's operation,

\begin{equation}
\begin{array}{lcl}
|\psi_{2}^{e'}\rangle & = & {\rm E_{C_{2}}}\left[{\rm E_{C_{1}}}\left({\rm |0\rangle_{A}|\phi^{+}\rangle_{{\rm C_{1}C_{2}}}|1\rangle_{{\rm B}}}\right)|\zeta\rangle\right]|\eta\rangle\\
 & = & {\rm E_{C_{2}}}\left[{\rm E_{C_{1}}}\frac{1}{\sqrt{2}}\left(|0\rangle|00\rangle|1\rangle+|0\rangle|11\rangle|1\rangle\right)_{{\rm AC_{1}C_{2}B}}|\zeta\rangle\right]|\eta\rangle\\
 & = & \frac{1}{\sqrt{2}}\left[\left({\rm A_{\zeta}A_{\eta}}|0\rangle|00\rangle|1\rangle|\zeta_{00}\rangle|\eta_{00}\rangle+{\rm A_{\zeta}B_{\eta}}|0\rangle|01\rangle|0\rangle|\zeta_{00}\rangle|\eta_{01}\rangle\right.\right.\\
 & + & \left.{\rm B_{\zeta}A_{\eta}}|1\rangle|10\rangle|1\rangle|\zeta_{01}\rangle|\eta_{00}\rangle+{\rm B_{\zeta}B_{\eta}}|1\rangle|11\rangle|0\rangle|\zeta_{01}\rangle|\eta_{01}\rangle\right)\\
 & + & \left({\rm A_{\zeta}A_{\eta}}|1\rangle|11\rangle|0\rangle|\zeta_{11}\rangle|\eta_{11}\rangle+{\rm A_{\zeta}B_{\eta}}|1\rangle|10\rangle|1\rangle|\zeta_{11}\rangle|\eta_{10}\rangle\right.\\
 & + & {\rm \left.\left.B_{\zeta}A_{\eta}|0\rangle|01\rangle|0\rangle|\zeta_{10}\rangle|\eta_{11}\rangle+{\rm B_{\zeta}B_{\eta}}|0\rangle|00\rangle|1\rangle|\zeta_{10}\rangle|\eta_{10}\rangle\right)_{AC_{1}C_{2}B}\right]}.
\end{array}\label{eq:state after Eve's operation}
\end{equation}

Alice and Bob receive their corresponding qubits from Charlie after
Eve's operation on the channel qubit and her \emph{ancilla }qubit. Then, they perform the CNOT operation accordingly to the protocol
and do the Bell measurements on their two-qubit system. The following
equation describes the above operations and possible measurement results
to be obtained by Alice and Bob,

\begin{equation}
\begin{array}{lcl}
|\psi_{2}^{e}\rangle & = & {\rm CNOT_{C_{1}\rightarrow A}CNOT_{C_{2}\rightarrow B}}|\psi_{2}^{e'}\rangle\\
 & = & \frac{1}{2\sqrt{2}}\left[\left\{ {\rm A_{\zeta}A_{\eta}\left(|\phi^{+}\rangle|\psi^{+}\rangle+|\phi^{+}\rangle|\psi^{-}\rangle+|\phi^{-}\rangle|\psi^{+}\rangle+|\phi^{-}\rangle|\psi^{-}\rangle\right)_{AC_{1}C_{2}B}|\zeta_{00}\rangle|\eta_{00}}\rangle\right.\right.\\
 & + & {\rm A_{\zeta}B_{\eta}\left(|\phi^{+}\rangle|\psi^{+}\rangle-|\phi^{+}\rangle|\psi^{-}\rangle+|\phi^{-}\rangle|\psi^{+}\rangle-|\phi^{-}\rangle|\psi^{-}\rangle\right)_{AC_{1}C_{2}B}|\zeta_{00}\rangle|\eta_{01}}\rangle\\
 & + & {\rm B_{\zeta}A_{\eta}}\left(|\phi^{+}\rangle|\psi^{+}\rangle+|\phi^{+}\rangle|\psi^{-}\rangle-|\phi^{-}\rangle|\psi^{+}\rangle-|\phi^{-}\rangle|\psi^{-}\rangle\right)_{{\rm AC_{1}C_{2}B}}|\zeta_{01}\rangle|\eta_{00}\rangle\\
 & + & \left.{\rm B_{\zeta}B_{\eta}}\left(|\phi^{+}\rangle|\psi^{+}\rangle-|\phi^{+}\rangle|\psi^{-}\rangle-|\phi^{-}\rangle|\psi^{+}\rangle+|\phi^{-}\rangle|\psi^{-}\rangle\right)_{{\rm AC_{{\rm 1}}C_{2}B}}|\zeta_{01}\rangle|\eta_{01}\rangle\right\} \\
 & + & \left\{ {\rm A_{\zeta}A_{\eta}}\left(|\phi^{+}\rangle|\psi^{+}\rangle-|\phi^{+}\rangle|\psi^{-}\rangle-|\phi^{-}\rangle|\psi^{+}\rangle+|\phi^{-}\rangle|\psi^{-}\rangle\right)_{{\rm AC_{1}C_{2}B}}|\zeta_{11}\rangle|\eta_{11}\rangle\right.\\
 & + & {\rm A_{\zeta}B_{\eta}}\left(|\phi^{+}\rangle|\psi^{+}\rangle+|\phi^{+}\rangle|\psi^{-}\rangle-|\phi^{-}\rangle|\psi^{+}\rangle-|\phi^{-}\rangle|\psi^{-}\rangle\right)_{{\rm AC_{1}C_{2}B}}|\zeta_{11}\rangle|\eta_{10}\rangle\\
 & + & {\rm B_{\zeta}A_{\eta}}\left(|\phi^{+}\rangle|\psi^{+}\rangle-|\phi^{+}\rangle|\psi^{-}\rangle+|\phi^{-}\rangle|\psi^{+}\rangle-|\phi^{-}\rangle|\psi^{-}\rangle\right)_{{\rm AC_{1}C_{2}B}}|\zeta_{10}\rangle|\eta_{11}\rangle\\
 & + & \left.\left.{\rm B_{\zeta}B_{\eta}}\left(|\phi^{+}\rangle|\psi^{+}\rangle+|\phi^{+}\rangle|\psi^{-}\rangle+|\phi^{-}\rangle|\psi^{+}\rangle+|\phi^{-}\rangle|\psi^{-}\rangle\right)_{{\rm AC_{1}C_{2}B}}|\zeta_{10}\rangle|\eta_{10}\rangle\right\} \right].
\end{array}\label{eq:after CNOT and Bell mesurement by Alice and Bob}
\end{equation}

Eve will be detected if Alice and Bob's measurement results are $|\phi^{+}\rangle$ and $|\psi^{-}\rangle$, or $|\phi^{-}\rangle$
and $|\psi^{+}\rangle$, respectively, after performing the Bell measurement. In such cases, the probability of detecting Eve can be expressed as:

\begin{equation}
\begin{array}{lcl}
{\rm P_{d}^{2}} & {\rm =} & {\rm \frac{1}{2}\left[A_{\zeta}^{2}A_{\eta}^{2}\left(1+\cos\alpha_{\zeta}\cos\alpha_{\eta}\right)+A_{\zeta}^{2}B_{\eta}^{2}\left(1+\cos\alpha_{\zeta}\cos\beta_{\eta}\right)\right.}\\
 & + & {\rm \left.B_{\zeta}^{2}A_{\eta}^{2}\left(1+\cos\beta_{\zeta}\cos\alpha_{\eta}\right)+B_{\zeta}^{2}B_{\eta}^{2}\left(1+\cos\beta_{\zeta}\cos\beta_{\eta}\right)\right],}
\end{array}\label{eq:Detection probability of second state}
\end{equation}
here, ${\rm P_{d}^{2}}$
is the detection probability of Eve's presence with the expected outcome
defined by the Eq$.$ (\ref{eq:=00005CPsi2}), and the superscript
$i$ (here $2$) represents as $i^{th}$ (here $2$nd) combination
among different combinations of chosen state by Alice, Bob, and Charlie
with probability ${\rm P}_{{\rm d}}^{i}$, $i\in\{1,2,\cdots,8\}$.
We may also calculate the detection probability of Eve for the other
outcomes, and get the same values i.e., ${\rm P_{d}^{1}=P_{d}^{2}=\cdots=P_{d}^{8}}$.
Combining with these results, one can easily evaluate the average
detection probability of Eve's presence as,

\begin{equation}
\begin{array}{lcl}
{\rm P_{d}} & = & {\rm \frac{1}{8}}\sum_{i}{\rm P}_{{\rm d}}^{i}\\
 & = & \frac{1}{2}\left[A_{\zeta}^{2}A_{\eta}^{2}\left(1+\cos\alpha_{\zeta}\cos\alpha_{\eta}\right)+A_{\zeta}^{2}B_{\eta}^{2}\left(1+\cos\alpha_{\zeta}\cos\beta_{\eta}\right)\right.\\
 & + & {\rm \left.B_{\zeta}^{2}A_{\eta}^{2}\left(1+\cos\beta_{\zeta}\cos\alpha_{\eta}\right)+B_{\zeta}^{2}B_{\eta}^{2}\left(1+\cos\beta_{\zeta}\cos\beta_{\eta}\right)\right].}
\end{array}\label{eq:average detection probabilty}
\end{equation}

\begin{figure}[h]
\begin{centering}
\includegraphics[scale=0.5]{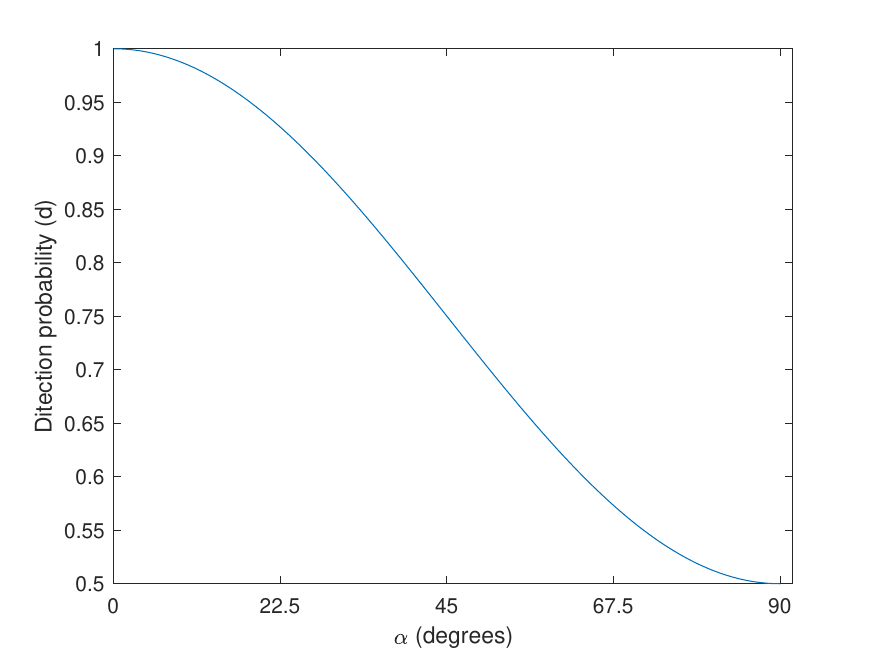}
\par\end{centering}
\caption{\label{fig:The-relationship-between d =000026 Alpha}(Color online)
The relationship between the detection probability of Eve $({\rm d})$
and the angle to describe non-orthogonality $({\rm \alpha})$.}
\end{figure}

Now, we consider the fact that Eve should adopt a strategy to reduce
the detection possibility of Eve can be as lower as possible. The
optimal scenario for Eve can be obtained with the condition of ${\rm A_{\zeta}=A_{\eta}=1}.$
Further, we impose another condition for \emph{optimal Eve incoherent
attack consists in a balanced one }considering $\alpha_{\zeta}=\alpha_{\eta}=\alpha$
\citep{LM_05}. We represent ${\rm d}$ the \emph{minimal detection
probability} of Eve,

\begin{equation}
\begin{array}{lcl}
{\rm d}\equiv{\rm min(P_{d})} & = & \frac{1}{2}\left(1+\cos^{2}\alpha\right).\end{array}\label{eq:min detection probability}
\end{equation}
The relation of the Eq. (\ref{eq:min detection probability}) is shown
in Figure \ref{fig:The-relationship-between d =000026 Alpha}, and
it can be noted that the \emph{minimal detection probability} of Eve
will be zero if she uses the orthogonal state\footnote{Here, we consider the optimal condition for Eve to avoid the rigorous
calculation. One can further proceed with the generalized assumption
(i.e., ${\rm A_{\zeta}^{2}+B_{\zeta}^{2}=1}$, ${\rm A_{\eta}^{2}+B_{\eta}^{2}=1}$,
$\alpha_{\zeta}\neq\alpha_{\eta}$ and $\beta_{\zeta}\neq\beta_{\eta}$)
to explore the dependence of the relation between \emph{minimal detection
probability} (d) of Eve and non-orthogonality of Eve's state. }. 

Next, we investigate Eve's information on the final one-bit key. Eve's attacking operations ${\rm E_{C_{1}}}$ and ${\rm E_{C_{2}}}$ on the qubits in the channels, i.e., CA channel and CB channel consist of the attack strategy $\mathcal{E}$. So, Eve's information on the one-bit key information $\mathcal{K}$ under the attack strategy $\mathcal{E}$ can be calculated as follows,

\begin{equation}
I\left(\mathcal{K}:\mathcal{E}\right)=H\left(\mathcal{E}\right)-H\left(\mathcal{E}|\mathcal{K}\right),\label{eq:Eve's information}
\end{equation}
where $H\left(\mathcal{E}\right)$ is the entropy of Eve's attack
strategy, and $H\left(\mathcal{E}|\mathcal{K}\right)$ is the conditional
entropy. Applying the optimal situation for Eve which is already mentioned,
the Eq$.$ (\ref{eq:after CNOT and Bell mesurement by Alice and Bob})
can be simplified as,

\begin{equation}
\begin{array}{lcl}
|\psi_{2}^{e}\rangle & = & \frac{1}{2\sqrt{2}}\left[\left\{ {\rm \left(|\phi^{+}\rangle|\psi^{+}\rangle+|\phi^{+}\rangle|\psi^{-}\rangle+|\phi^{-}\rangle|\psi^{+}\rangle+|\phi^{-}\rangle|\psi^{-}\rangle\right)_{{\rm AC_{1}C_{2}B}}|\zeta_{00}\rangle|\eta_{00}}\rangle\right\} \right.\\
 & + & \left.\left\{ \left(|\phi^{+}\rangle|\psi^{+}\rangle-|\phi^{+}\rangle|\psi^{-}\rangle-|\phi^{-}\rangle|\psi^{+}\rangle+|\phi^{-}\rangle|\psi^{-}\rangle\right)_{{\rm AC_{1}C_{2}B}}|\zeta_{11}\rangle|\eta_{11}\rangle\right\} \right].
\end{array}\label{eq:Optimal Eve's detection}
\end{equation}

The above equation shows that Eve's measurement result is either $|\zeta_{00}\rangle|\eta_{00}\rangle$
or $|\zeta_{11}\rangle|\eta_{11}\rangle$ with equal possibility $\frac{1}{2}$.
If we again consider the different combinations of preparation states
by Charlie, Alice, and Bob (i.e., $|\psi_{1}^{e}\rangle,|\psi_{3}^{e}\rangle,\cdots,|\psi_{8}^{e}\rangle$),
Eve will have one of the two measurement results with probability
$\frac{1}{2}$ by the attack strategy $\mathcal{E}$. We get the Shannon
entropy of $\mathcal{E}$, i.e., $H\left(\mathcal{E}\right)=1$, and the
conditional entropy of $\mathcal{E}$ when given $\mathcal{K}$ is
$H\left(\mathcal{E}|\mathcal{K}\right)=h_{2}\left(Q_{\mathcal{E\mathcal{K}}}\right)$,
here $h_{2}$ is the binary entropy function, and $Q_{\mathcal{E}\mathcal{K}}$
is the quantum bit error rate (QBER) introduced by Eve in an effort
to predict the final one-bit key after using the attack strategy $\mathcal{E}$.
Now, we have to calculate the expression for $Q_{\mathcal{E}\mathcal{K}}$.
The possibility of correctly distinguishing between two quantum states
with scalar product $\cos\alpha$ is $\frac{1}{2}\left(1+\sin\alpha\right)$
\citep{H1969,GRT+02}, Eve may guess the final key if she can distinguish
correctly the states $\zeta$ and $\eta$, and  there is also another possibility
that Eve makes a mistake in distinguishing the state $\zeta$ as well
as $\eta$. Here, one error would compel Eve to misinterpret
the final key but the second error compensates for the first, eventually,
Eve will able to guess the final key correctly. The general expression
for $Q_{\mathcal{E}\mathcal{K}}$ in terms of $\alpha_{\zeta}$ and
$\alpha_{\eta}$ is as follows,

\begin{equation}
\begin{array}{lcl}
Q_{\mathcal{E}\mathcal{K}} & =1- & \left[\left(\frac{1+\sin\alpha_{\zeta}}{2}\right)\left(\frac{1+\sin\alpha_{\eta}}{2}\right)+\left(\frac{1-\sin\alpha_{\zeta}}{2}\right)\left(\frac{1-\sin\alpha_{\eta}}{2}\right)\right]\\
 & = & \frac{1-\sin\alpha_{\zeta}\sin\alpha_{\eta}}{2},
\end{array}\label{eq:QBER expression}
\end{equation}
now, using the optimal condition for Eve (i.e., $\alpha_{\zeta}=\alpha_{\eta}=\alpha$)
with Eq$.$ (\ref{eq:Eve's information}) and Eq$.$ (\ref{eq:min detection probability}),
we have

\begin{equation}
\begin{array}{lcl}
I\left(\mathcal{K}:\mathcal{E}\right) & = & \frac{1}{2}\left[1-h_{2}\left(\frac{1-\sin^{2}\alpha}{2}\right)\right],\end{array}\label{eq:final expresion for Eve's information}
\end{equation}
where the $\frac{1}{2}$ factor comes due to the fact that in Eq. (\ref{eq:Optimal Eve's detection}),
the outcome of Alice and Bob are the same and equally probable irrespective
of Eve's outcome.

\begin{figure}[h]
\begin{centering}
\includegraphics[scale=0.5]{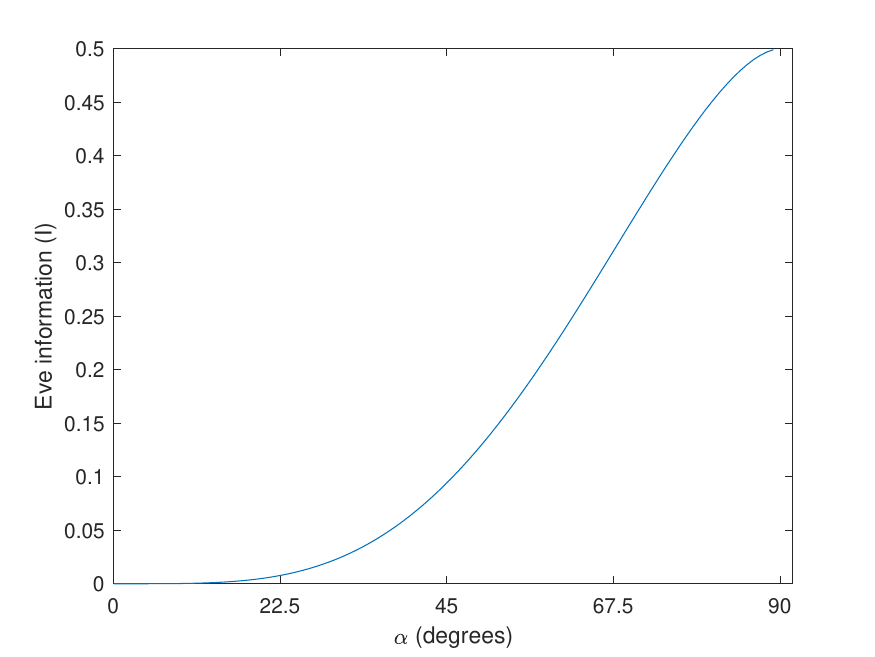}
\par\end{centering}
\caption{\label{fig:The-relationship-between I =000026 Alpha}(Color online)
The relationship between Eve's information $({\rm I})$ and angle
to describe non-orthogonality $({\rm \alpha})$.}

\end{figure}

The relationship between $I$ and $\alpha$ is illustrated in Figure
\ref{fig:The-relationship-between I =000026 Alpha} for the optimal
situation of Eve. One can see from Figure \ref{fig:The-relationship-between I =000026 Alpha},
if Eve uses orthogonal states (i.e., $\alpha=\frac{\pi}{2}$), the
maximum information that can be gained by Eve will be $0.5$ bit.
Detection probability of Eve will be zero, but if she takes non-orthogonal
states the information gain will not be maximum in any case\footnote{It may be noted that this conclusion will not be the same if we do
not consider $B_{\eta}=B_{\zeta}=0$ and $A_{\eta}=A_{\zeta}=1$.
One can do further analysis with the generalized assumption.}.  It is important to note that in real communication
scenarios, there exists a probability of introducing noise when Eve
performs a unitary operation with the ancilla state, denoted as $A_{\eta}\,{\rm and}\,A_{\zeta}<1$.
Consequently, executing this attack becomes challenging for Eve in
optimal conditions. Under these circumstances, Eve cannot remain undetected,
even when utilizing orthogonal basis states. Additionally, in the
quantum systems' transfer among legitimate parties, the use of decoy
states is a common practice to enhance the security of the quantum
channel (see step-wise description of Protocol 1). It is noteworthy
that we have not considered decoy states in our security analysis
for the sake of simplicity in our security description. If one takes
into account the security provided by decoy states, the detection
probability will be higher compared to not considering decoy states.
In the subsequent section, we calculate the tolerable error bound,
demonstrating that our protocol remains secure under collective attack. 

From Table \ref{tab:Relation of final key and announcement for P1},
we can conclude that the final agreement key after key execution of
the Protocol 1 is maximally uncertain or the entropy is maximum due
to classical announcement of our protocol, and further we want to
stress on the fact that the assumption for simplification of our security
derivation leads to symmetric output of Alice and Bob's measurement
(see Eq. (\ref{eq:Optimal Eve's detection})) with respect to the measurement
outcome of Eve's ancillary state. Despite knowing the fact, we evaluate
the success probability of estimating the final key by Eve in a generalize
manner\footnote{Depending upon the protocol and Eve's collective attack, the probability
of successful estimation of the final key by Eve can be expressed
as the detection probability of Eve rather than a constant value.
For that reason, we use the more generalized approach.}. 

According to Eve's measurement outcomes $|\zeta_{ij}\rangle|\eta_{kl}\rangle$ with $i,j,k,l\in\{0,1\}$, and Charlie's announcement, Eve can decide the final one-bit key information. For generalization, we can consider that Eve decides the key to be $0$ with probability $e$ and $1$ with probability\footnote{In our protocol, the final key appears symmetrically in respect of Charlie's announcement. So, in our case, the contribution of $e$ is  worth noting.} $1-e$. Now, we are interested to calculate the total probability of Eve's successfully estimating the final key value,

\begin{equation}
\begin{array}{lcl}
{\rm Pr_{s}} & = & \left(\frac{1+\sin\alpha}{2}\right)^{2}\left\{ \frac{1}{2}e+\frac{1}{2}(1-e)\right\} +\left(\frac{1-\sin\alpha}{2}\right)^{2}\left\{ \frac{1}{2}e+\frac{1}{2}(1-e)\right\} \\
 & + & \frac{\left(1+\sin\alpha\right)\left(1-\sin\alpha\right)}{2}\left\{ \frac{1}{2}e+\frac{1}{2}(1-e)\right\} \\
 & = & \frac{1}{2}.
\end{array}\label{eq:Eve's successfully guessing the two-bit key}
\end{equation}

Using Eq. (\ref{eq:min detection probability}) and Eq. (\ref{eq:Eve's successfully guessing the two-bit key}),
we get the probability of Eve's successfully obtaining the final key
with the value ${\rm n}$ i.e., the number of the final key is,

\begin{equation}
\begin{array}{lcl}
{\rm Pr} & = & {\rm \left[Pr_{s}\left(1-d\right)\right]^{n}}\\
 & = & {\rm \left[\frac{1}{2}\left(1-d\right)\right]^{n}.}
\end{array}\label{eq:Eve's=000020success=000020probability}
\end{equation}

\begin{figure}[h]
\begin{centering}
\includegraphics[scale=0.5]{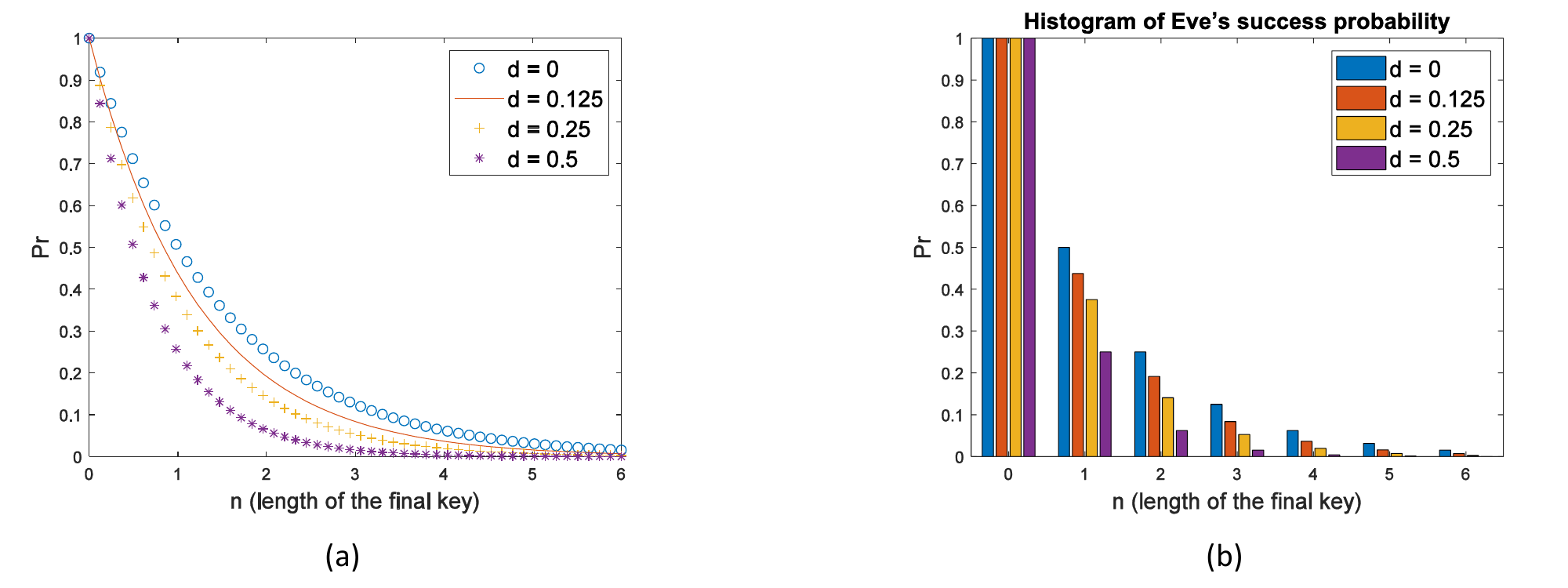}
\par\end{centering}
\caption{\protect\label{fig:P vs n and d}(Color online) The relationship among ${\rm Pr,d}$ and ${\rm n}$ is illustrated.
(a) The variation of ${\rm Pr}$ with ${\rm n}$ for different values
of ${\rm d}$ is shown. Specifically, ${\rm d=0\%}$ for the ``circle''
line, ${\rm d=12.5\%}$ for the ``continuous'' line, ${\rm d=25\%}$
for ``plus sign'' line and ${\rm d=50\%}$ for ``asterisk'' line.
(b) A histogram of Eve's success probability for ${\rm n=6}$ is displayed
for different values of ${\rm d}$. }

\end{figure}

We have illustrated the relationship between ${\rm Pr}$ and ${\rm n}$
with different values of ${\rm d}$ in Figure \ref{fig:P vs n and d}.
Our plot shows that ${\rm Pr}$ is approximately zero for the ${\rm n=6}$;
for example, when the value of detection possibility of Eve (${\rm d}$)
is $25\%$, then the probability of Eve's successfully obtaining the
final key is $2.629\times10^{-3}$, i.e., it can be neglected. We present Figure \ref{fig:P vs n and d}.a
to determine the value of $n$ at which ${\rm Pr}$ approaches zero.
Since $n$ is an integer, we provide Figure \ref{fig:P vs n and d}.b,
which displays a histogram of ${\rm Pr}$ for discrete integer values
of $n$. Clearly,
we can state that when the ${\rm n}$ value is large enough the information
achieved by Eve with impersonated fraudulent attack can be eliminated.

\emph{Tolerable error bound under collective attack
scenario}: Owing to the symmetry in the outcomes of the composite system, we consider
the state $|\psi_{2}^{e}\rangle$, shared by Alice, Bob, and Eve is
of the form described in Eq. (\ref{eq:Optimal Eve's detection}). In
the case of collective attack \citep{PAB+09,DW05}, the lower bound
for the secret key rate in the asymptotic limit is determined by the
Devetak-Winter rate equation. This equation expresses the difference
between the mutual information shared between Alice and Bob and the
Holevo quantity between Eve and Bob\footnote{In this context, we evaluate the Holevo bound between
Eve and Bob. The classical one-way postprocessing occurs from Bob
to Alice, and it is directly linked to his outcome after the Bell
measurement.},

\begin{equation}
r\ge r_{{\rm DW}}=I\left(A:B\right)-\chi\left(B:E\right).\label{eq:Devetak-Winter Key rate}
\end{equation}
Here, $r$ is secret key rate in asymptotic limit and $r_{DW}$ is
the lower bound of $r$, given by Devetak-Winter rate equation.
In the present context, the information measure \citep{PAB+09}  $I\left(A:B\right)=H\left(B\right)-H\left(B|A\right)$
and $\chi\left(B:E\right)=S\left(\rho_{E}\right)-\underset{k_{B}=0,1}{\sum}p_{k_{B}}S\left(\rho_{E|k_{B}}\right)$.
Here, $S$ represents the standard von Neumann entropy, $\rho_{E}={\rm Tr}_{AB}|\psi_{2}^{e}\rangle\langle\psi_{2}^{e}|$
denotes Eve's quantum system after tracing out Alice's and Bob's quantum
system, and $\rho_{E|k_{B}}$ represents Eve's quantum system when
Bob obtains the measurement result $k_{B}$. The probabilities of
Bob's final results are equal, i.e., $p_{0}=p_{1}=\frac{1}{2}$ after
considering error. Assuming uniformity in the error rates for measurements
in both Alice and Bob's measurement results, let us denote this common
error rate as $\epsilon$. The expression may be derived $I\left(A:B\right)=1-h\left(\epsilon\right)$.
Utilizing Eq. (\ref{eq:=00005CPsi2}), we can present Eq. (\ref{eq:Optimal Eve's detection})
in a generalized form\footnote{We exclude the consideration of $\eta$ ancilla states,
as they do not provide additional information under the conditions
$B_{\eta}=B_{\zeta}=0$ and $A_{\eta}=A_{\zeta}=1$, which represent
the optimal scenario for Eve.} by considering the error probability $\epsilon$
which accounts for potential errors introduced by both the quantum
channel and eavesdropping as \citep{LF+11},

\begin{equation}
\begin{array}{lcl}
|\psi_{2}^{e}\rangle & = & \sqrt{\frac{1-\epsilon}{4}}|\phi^{+}\rangle_{{\rm AC_{1}}}|\psi^{+}\rangle_{{\rm C}_{2}B}\left(|\zeta_{00}\rangle+|\zeta_{11}\rangle\right)+\sqrt{\frac{\epsilon}{4}}|\phi^{+}\rangle_{{\rm AC_{1}}}|\psi^{-}\rangle_{{\rm C_{2}B}}\left(|\zeta_{00}\rangle-|\zeta_{11}\rangle\right)\\
 & + & \sqrt{\frac{\epsilon}{4}}|\phi^{-}\rangle_{{\rm AC_{1}}}|\psi^{+}\rangle_{{\rm C_{2}B}}\left(|\zeta_{00}\rangle-|\zeta_{11}\rangle\right)+\sqrt{\frac{1-\epsilon}{4}}|\phi^{-}\rangle_{{\rm AC_{1}}}|\psi^{-}\rangle_{{\rm C_{2}B}}\left(|\zeta_{00}\rangle+|\zeta_{11}\rangle\right)
\end{array}.\label{eq:Composite state with error probability}
\end{equation}

After conducting measurements in the Bell basis on the systems associated
with Alice and Bob, it would be straightforward to confirm the state
of Eve's system which may be denoted by $|\theta^{\phi^{\pm},k_{B}}\rangle$,
where $\phi^{\pm}$ and $k_{B}$ represent the outcomes of Alice and
Bob, respectively. The explicit form of Eve's state is as follows:

\begin{equation}
\begin{array}{lcl}
|\theta^{\phi^{+},1}\rangle & = & \sqrt{\frac{1-\epsilon}{4}}\left(|\zeta_{00}\rangle+|\zeta_{11}\rangle\right)\\
|\theta^{\phi^{+},0}\rangle & = & \sqrt{\frac{\epsilon}{4}}\left(|\zeta_{00}\rangle-|\zeta_{11}\rangle\right)\\
|\theta^{\phi^{-},1}\rangle & = & \sqrt{\frac{\epsilon}{4}}\left(|\zeta_{00}\rangle-|\zeta_{11}\rangle\right)\\
|\theta^{\phi^{-},0}\rangle & = & \sqrt{\frac{1-\epsilon}{4}}\left(|\zeta_{00}\rangle+|\zeta_{11}\rangle\right)
\end{array}.\label{eq:Eve's state after Alice's and Bob's trace out}
\end{equation}
We express Eve's system in the density operator formalism, with a
conditional dependence on the output of Bob,

\begin{equation}
\begin{array}{lcl}
\sigma_{E}^{0}=\rho_{E|0} & = & \epsilon|\theta^{\phi^{+},0}\rangle\langle\theta^{\phi^{+},0}|+\left(1-\epsilon\right)|\theta^{\phi^{-},0}\rangle\langle\theta^{\phi^{-},0}|\\
 & = & \left(\frac{\epsilon^{2}}{4}+\frac{\left(1-\epsilon\right)^{2}}{4}\right)\left[|\zeta_{00}\rangle\langle\zeta_{00}|+|\zeta_{11}\rangle\langle\zeta_{11}|\right]+\left(-\frac{\epsilon^{2}}{4}+\frac{\left(1-\epsilon\right)^{2}}{4}\right)\left[|\zeta_{00}\rangle\langle\zeta_{11}|+|\zeta_{11}\rangle\langle\zeta_{00}|\right],\\
\\\sigma_{E}^{1}=\rho_{E|1} & = & \left(1-\epsilon\right)|\theta^{\phi^{+},1}\rangle\langle\theta^{\phi^{+},1}|+\epsilon|\theta^{\phi^{-},1}\rangle\langle\theta^{\phi^{-},1}|\\
 & = & \left(\frac{\epsilon^{2}}{4}+\frac{\left(1-\epsilon\right)^{2}}{4}\right)\left[|\zeta_{00}\rangle\langle\zeta_{00}|+|\zeta_{11}\rangle\langle\zeta_{11}|\right]+\left(-\frac{\epsilon^{2}}{4}+\frac{\left(1-\epsilon\right)^{2}}{4}\right)\left[|\zeta_{00}\rangle\langle\zeta_{11}|+|\zeta_{11}\rangle\langle\zeta_{00}|\right],
\end{array}\label{eq:Conditionl density matrix of Eve's system}
\end{equation}
and after tracing out the systems belonging to Alice and Bob, the
density matrix of Eve's system can be obtained as

\begin{equation}
\rho_{E}=\frac{1}{2}\left[\left(|\zeta_{00}\rangle\langle\zeta_{00}|+|\zeta_{11}\rangle\langle\zeta_{11}|\right)+\left(1-2\epsilon\right)\left(|\zeta_{00}\rangle\langle\zeta_{11}|+|\zeta_{11}\rangle\langle\zeta_{00}|\right)\right].\label{eq:Density matrix of Eve's system after tracing out Alice's and Bob's system}
\end{equation}
The secret key rate can now be expressed utilizing Eqs. (\ref{eq:Devetak-Winter Key rate}),
(\ref{eq:Conditionl density matrix of Eve's system}), and (\ref{eq:Density matrix of Eve's system after tracing out Alice's and Bob's system})
as 

\begin{equation}
\begin{array}{lcl}
r\ge r_{{\rm DW}} & = & 1-h\left(\epsilon\right)-\left[S\left(\rho_{E}\right)-\left(\frac{1}{2}S\left(\sigma_{E}^{0}\right)+\frac{1}{2}S\left(\sigma_{E}^{1}\right)\right)\right]\end{array}.\label{eq:Final key rate equation}
\end{equation}
To calculate $r_{DW}$, we commence
by selecting an orthogonal basis $|E_{00}\rangle,|E_{01}\rangle,|E_{10}\rangle,$ and
$|E_{11}\rangle$ within the Hilbert space $\mathscr{H}^{E}$. Consequently,
we consider the states in the orthogonal basis $|\zeta_{00}\rangle=\sum_{ij}a_{ij}E_{ij}$
and $|\zeta_{11}\rangle=\sum_{ij}f_{ij}E_{ij}$, where $i,j\in\{0,1\}$.
This is straight forward to have these constraints $|a_{00}|^{2}+|a_{01}|^{2}+|a_{10}|^{2}+|a_{11}|^{2}=1$,
$|f_{00}|^{2}+|f_{01}|^{2}+|f_{10}|^{2}+|f_{11}|^{2}=1$, and $a_{00}^{*}f_{00}+a_{01}^{*}f_{01}+a_{10}^{*}f_{10}+a_{11}^{*}f_{11}=\cos\alpha$.
Following a detailed computation, the eigenvalues of $\rho_{E}$ are
determined as $\lambda_{1,2}^{\rho_{E}}=0,0$, $\lambda_{3}^{\rho_{E}}=\frac{1}{2}\left(1+{\rm cos}\alpha-2\epsilon{\rm cos}\alpha-\sqrt{\left(1-2\epsilon+{\rm cos}\alpha\right)^{2}}\right)$
and $\lambda_{4}^{\rho_{E}}=\frac{1}{2}\left(1+{\rm cos}\alpha-2\epsilon{\rm cos}\alpha+\sqrt{\left(1-2\epsilon+{\rm cos}\alpha\right)^{2}}\right)$.
Similarly, the eigenvalues of $\sigma_{E}^{1}$ or $\sigma_{E}^{0}$
are computed as $\lambda_{1,2}^{\sigma_{E}^{\pm}}=0,0$, $\lambda_{3}^{\sigma_{E}^{\pm}}=\frac{1}{4}\left(1-2\epsilon+2\epsilon^{2}+{\rm cos}\alpha-2\epsilon{\rm cos}\alpha-\sqrt{\left(1-2\epsilon+\left(1-2\epsilon+2\epsilon^{2}\right){\rm cos}\alpha\right)^{2}}\right)$
and\\
{} $\lambda_{4}^{\sigma_{E}^{\pm}}=\frac{1}{4}\left(1-2\epsilon+2\epsilon^{2}+{\rm cos}\alpha-2\epsilon{\rm cos}\alpha+\sqrt{\left(1-2\epsilon+\left(1-2\epsilon+2\epsilon^{2}\right){\rm cos}\alpha\right)^{2}}\right)$.
Subsequently, the entropy relations are employed as $S\left(\rho_{E}\right)=-\sum_{i}\lambda_{i}^{\rho_{E}}{\rm log_{2}\lambda_{i}^{\rho_{E}}}$
and $S\left(\sigma_{E}^{\pm}\right)=-\sum_{i}\lambda_{i}^{\sigma_{E}^{\pm}}{\rm log_{2}\lambda_{i}^{\sigma_{E}^{\pm}}}$
to yield the explicit form of Eq. (\ref{eq:Final key rate equation}). 

\begin{figure}[h]
\begin{centering}
\includegraphics[scale=0.5]{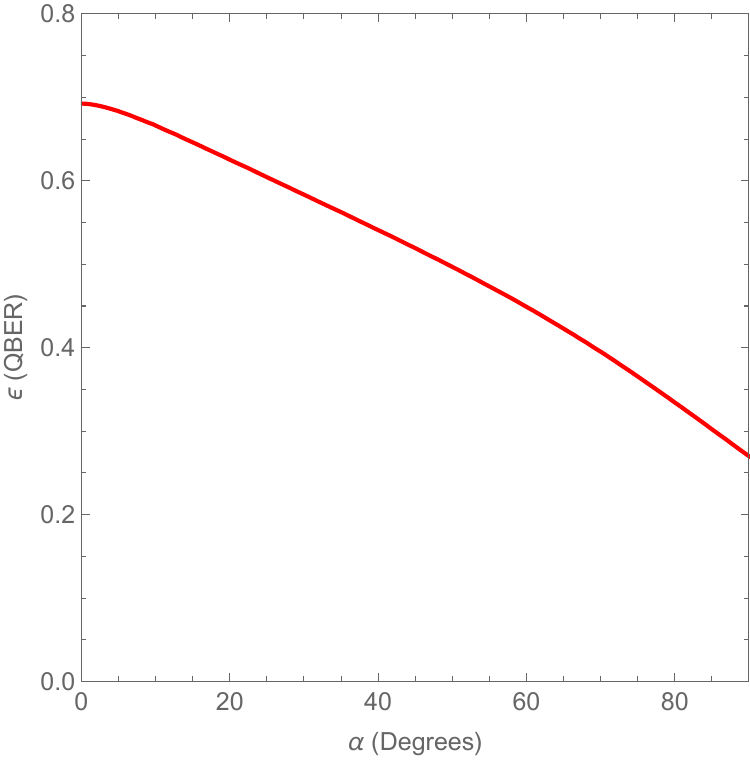}
\par\end{centering}
\caption{\label{fig:QBER vs Alpha}(Color online) Plot shows
the variation of tolerable QBER ($\epsilon$) with the angle to describe
non-orthogonality $({\rm \alpha})$.}

\end{figure}

To determine the tolerable error limit under secure
bound, commonly referred to as tolerable QBER, we employ the solution
of Eq. (\ref{eq:Final key rate equation}) involving the variable
$\alpha$. In order to visually represent this observation, we generate
a plot illustrating the tolerable error probability, $\epsilon$,
as a function of the angle $\alpha$, which characterizes non-orthogonality
(cf. Figure \ref{fig:QBER vs Alpha}). It is evident that an increase
in the non-orthogonality of Eve's ancilla state corresponds to an
augmented value of $\epsilon$. Specifically, when Eve employs an
orthogonal state as the ancilla state ($\alpha=\frac{\pi}{2}$),
the permissible error threshold is determined to be $27\%$. This
limit serves as a robust indicator of the security of our protocol
in the context of a collective attack scenario. It
is noteworthy that when the threshold limit is set at $27\%$, the
corresponding value for $r$ is zero. However, it is observed that
for the same $\alpha$ value, as depicted in Figure \ref{fig:The-relationship-between d =000026 Alpha},
the error probability reaches 0.5. This phenomenon arises due to the
possibility of the user obtaining a non-zero value for $r$.

\section{Effect of noise on proposed CQKA protocol\protect\label{sec:Effect_of_Noise}}

Environmental-induced
decoherence \citep{walls1994atomic,BLP+16,ZG97} is a well-recognized
limitation to the efficiency of quantum-enhanced protocols \citep{XDS+13,TPB17}.
This phenomenon poses a significant challenge in implementing quantum
communication schemes. Even in the absence of an adversary, interactions
between qubits and their surrounding environment can compromise accuracy,
potentially leading to incorrect outcomes. Any practical realization
of these protocols is inherently susceptible to noise due to environmental
interference \citep{GC85}. For the protocol to be viable, it must
maintain correctness despite a limited level of noise. Moreover, when
the communication channel is exposed to an eavesdropper (Eve), she
could attempt to disguise her interference as environmental noise
\citep{ZG97,SWY22}. However, distinguishing between genuine noise
and malicious interference is possible using established detection
techniques, ensuring protocol integrity even under adversarial conditions
\citep{HTS22}.

In this work, we examine the impact of a noisy channel
on the protocols under consideration. Specifically, we analyze amplitude
damping and phase damping within a Markovian framework, as well as
dephasing and depolarizing interactions in the context of a non-Markovian
reservoir \citep{SXL+18,NBS24}. Notably, while non-Markovian dephasing
interactions can preserve entanglement for longer duration, entanglement
revival may also occur under dissipative interactions. The dynamics of the quantum system are described using the Kraus representation.
This analysis is crucial for understanding the system's behavior,
which is fundamental to implementing these two channels in quantum
information and communication.

The Kraus operators describing amplitude damping
(AD) are as follows \citep{nielsen2010quantum,MTP+22},

\begin{equation}
\begin{array}{ccccccccc}
\mathcal{F}_{0}^{{\rm AD}} & = & \left(\begin{array}{cc}
1 & 0\\
0 & \sqrt{1-\eta_{a}}
\end{array}\right) &  & {\rm and} &  & \mathcal{F}_{1}^{{\rm AD}} & = & \left(\begin{array}{cc}
1 & \sqrt{\eta_{a}}\\
0 & 0
\end{array}\right)\end{array},\label{eq:Amplitude_Damping}
\end{equation}
and for phase damping (PD) are

\begin{equation}
\begin{array}{ccccccccc}
\mathcal{F}_{0}^{{\rm PD}} & = & \left(\begin{array}{cc}
1 & 0\\
0 & \sqrt{1-\eta_{p}}
\end{array}\right) &  & {\rm and} &  & \mathcal{F}_{1}^{{\rm PD}} & = & \left(\begin{array}{cc}
1 & 0\\
0 & \sqrt{\eta_{p}}
\end{array}\right)\end{array},\label{eq:Phase_Damping}
\end{equation}
here, $\eta_{j},j\in\left\{ a,p\right\} $ is the damping parameter.
Similarly, the dynamics of non-Markovian dephasing (NMDPH) are characterized
by the Kraus operators given below \citep{NDB19}:

\begin{equation}
\begin{array}{lcl}
\mathcal{N}_{0}^{{\rm NMDPH}} & = & \sqrt{\left(1-\alpha p\right)\left(1-p\right)}\mathds{I},\\
\\\mathcal{N}_{0}^{{\rm NMDPH}} & = & \sqrt{p\left[1+\alpha\left(1-p\right)\right]}\sigma_{Z},
\end{array}\label{eq:DePhasing}
\end{equation}
and for non-Markovian depolarizing (NMDPO) channel \citep{SSB18},

\begin{equation}
\begin{array}{lcl}
\mathcal{N}_{\mathds{I}}^{{\rm NMDPO}} & = & \sqrt{\left(1-3\alpha p\right)\left(1-p\right)}\mathds{I},\\
\\\mathcal{N}_{X}^{{\rm NMDPO}} & = & \sqrt{\left[1+3\alpha\left(1-p\right)\right]\frac{p}{3}}\sigma_{X},\\
\\\mathcal{N}_{Y}^{{\rm NMDPO}} & = & \sqrt{\left[1+3\alpha\left(1-p\right)\right]\frac{p}{3}}\sigma_{Y},\\
\\\mathcal{N}_{Z}^{{\rm NMDPO}} & = & \sqrt{\left[1+3\alpha\left(1-p\right)\right]\frac{p}{3}}\sigma_{Z}.
\end{array}\label{eq:DePolarizing}
\end{equation}
The parameter $\alpha$ $\left(0\leq\alpha\le1\right)$ characterizes
the degree of non-Markovianity in the channel. Specifically, $\alpha=0$
represents standard dephasing, while higher values of $\alpha$ indicate
an increasing degree of non-Markovian behavior. Additionally, $p$
is a time-like parameter constrained by $0\leq p\le\frac{1}{2}$ \citep{NDB19},
which varies monotonically with time. Here, $\mathds{I}$ denotes
the identity matrix in a two-dimensional Hilbert space, and $\sigma_{X},\sigma_{Y}\,{\rm and}\,\sigma_{Z}$
are the Pauli matrices.

\begin{figure}[h]
\begin{centering}
\includegraphics[scale=0.5]{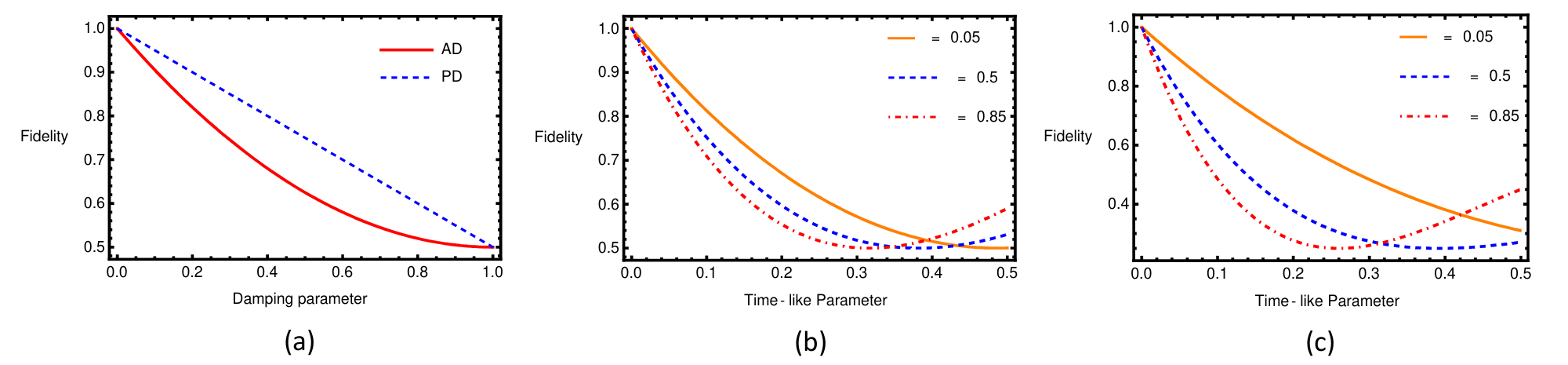}
\par\end{centering}
\caption{\protect\label{fig:NOISY_CHANNEL}(Color online) Variation of average
fidelity with noisy quantum channel. (a) Variation of average fidelity
for the CQKA scheme with respect to the channel parameter ($\eta_{j}$)
in the Markovian regime for AD and PD channels (as described by Eqs.
(\ref{eq:Amplitude_Damping}) and (\ref{eq:Phase_Damping})). (b)
and (c) Variation of average fidelity for NMDPH and NMDPO channels
in non-Markovian regime as a function of the time-like parameter $p$,
evaluated for different values of $\alpha$, respectively (see Eqs.
(\ref{eq:DePhasing}) and (\ref{eq:DePolarizing})).}

\end{figure}

Before analyzing the performance of noisy channels
on our scheme, it is crucial to highlight the initial setup. If an
$x$-qubit composite state (initial state), $\rho_{i}=|\Phi\rangle\langle\Phi|$,
is employed for implementing a protocol, where $y$ home qubits and
$x-y$ travel qubits are denoted as $h$ and $t$, respectively, the
general expression for the final state before measurement can be formulated
as:

\[
\begin{array}{lcl}
\rho_{f}^{k} & = & \underset{j}{\sum}\left\{ I_{h}^{\otimes y}\otimes\left(\mathcal{F}_{1}^{k}\otimes\ldots\otimes\mathcal{F}_{j}^{k}\otimes\ldots\otimes\mathcal{F}_{x-y}^{k}\right)_{t}\right\} \rho_{i}\left\{ I_{h}^{\otimes y}\otimes\left(\mathcal{F}_{1}^{k}\otimes\ldots\otimes\mathcal{F}_{j}^{k}\otimes\ldots\otimes\mathcal{F}_{x-y}^{k}\right)_{t}\right\} ^{\dagger},\end{array}
\]
where, $\mathcal{F}_{j}^{k}$ are the Kraus operators corresponding
to various noise channels, where $k\in\left\{ {\rm AD,PD,NMDPH,NMDPO}\right\} $.
The impact of noise on the protocol can be quantified using a distance-based
measure known as the square of fidelity (referred to as average fidelity)
\citep{nielsen2010quantum}. This average fidelity is expressed as:

\[
\begin{array}{lcl}
F^{k} & = & \langle\Phi^{f}|\rho_{f}^{k}|\Phi^{f}\rangle.\end{array}
\]
In this context, $|\Phi^{f}\rangle$ represents the final state that
the initial pure state $|\Phi\rangle$ would ideally attain after
all encoding operations are performed by each party in a decoherence-free
environment.

Now, we analyze the performance of the proposed scheme
considering the quantum channel as a noisy channel. Qubits ${\rm C_{1}}$
and ${\rm C_{2}}$ are transmitted through the noisy environment.
We first examine the Markovian regime, considering AD and PD channels.
To evaluate the average fidelity \citep{nielsen2010quantum} between
the state under noise-free conditions and the state affected by noise,
we plot the average fidelity as a function of the damping parameter
$\eta_{j}$ in Figure \ref{fig:NOISY_CHANNEL}.a. For the AD channel,
the fidelity decreases more rapidly with $\eta_{a}$ compared to the
PD channel. The fidelity expressions for the AD and PD channels are
given by $1-\frac{1}{2}\left(\eta-2\right)\eta$ and $1-\frac{\eta}{2}$,
respectively. Notably, the decrease in fidelity for the AD channel
is nonlinear, in contrast to the linear decrease observed for the
PD channel. At $\eta_{j}=1$, the fidelity reaches a minimum value
of $0.5$ for both cases, which represents the limit in a strong damping
environment.

In the non-Markovian regime, the time-like parameter
$p$ is parameterized as $p=\frac{1}{2}\left(1-e^{-\mathcal{K}t}\right)$,
with a range $0\leq p\le\frac{1}{2}$. The average fidelity for the
non-Markovian dephasing and depolarizing channels are given by $\frac{1}{2}\left[1+\left\{ 1-2p+2\left(p-1\right)p\,\alpha\right\} \right]$
and $1+\frac{2}{3}\left[3\left(p-1\right)\alpha-1\right]\left[6\left(p-1\right)p\,\alpha-2p+3\right]$,
respectively. Figure \ref{fig:NOISY_CHANNEL}.b illustrates the effect
of the non-Markovian dephasing channel on the proposed scheme. The
fidelity decreases as $p$ increases, with the rate of decline depending
on the value of $\alpha$. For lower $\alpha$ values, fidelity is
maintained over a longer duration compared to higher $\alpha$ values.
Notably, for larger $\alpha$, the fidelity exhibits a significant
revival, with the minimum value of fidelity reaching $0.5$. In Figure
\ref{fig:NOISY_CHANNEL}.c, a similar
trend is observed for the non-Markovian depolarizing channel. However,
the decrease in fidelity is more pronounced for the larger values
of $\alpha$ compared to the dephasing channel. Conversely, for $\alpha=0.05$,
the fidelity declines more slowly in the depolarizing channel than
in the dephasing channel. This indicates that fidelity is preserved
for a longer period in the depolarizing channel at lower $\alpha$
values. For higher $\alpha$ values $\left(\alpha=0.5,0.85\right)$,
fidelity is maintained for a shorter duration in the depolarizing
channel compared to the dephasing channel. Additionally, the non-Markovian
depolarizing channel exhibits a gradual improvement in fidelity after
a certain value of $p$, which depends on the specific $\alpha$ value.

It is important to consider a specific scenario:
as we will observe, leveraging non-Markovianity can lead to improved
performance (greater efficiency) compared to a Markovian environment.
Recognizing this, an eavesdropper (say, Eve) might substitute a Markovian
channel between Alice and Bob with a non-Markovian one to lower the
chances of detection (or just substitute a more noisy channel by a
less noisy one). Consequently, the legitimate parties must account
for this possibility when determining the tolerable error limit. Additionally,
setting this limit may require users to characterize the communication
channel. For instance, the noise in an amplitude damping channel could
be identified and mitigated accordingly \citep{OSB15}.

\section{Comparison of the proposed CQKA protocol with a set of existing protocols\label{sec:Comparison}}

In this section, we briefly compare our proposed CQKA scheme and QKA
scheme with similar kinds of schemes which are already been proposed
in the recent past. We intentionally avoid mentioning the multiparty-based
QKA schemes \citep{LZ+21} as our proposed protocol is restricted
to establishing the key agreement between two parties with and without
the help of a third party. A few parameters are involved to account
for the performance of QKA schemes, like quantum resources (QR), quantum efficiency (QE), quantum channel (QC) required, quantum memory (QM), presence of the third party (TP), number of parties (NoP) involved
to perform the protocol to accomplish key agreement task, and quantum
operation used by the legitimate parties. Our comparison is based
on the above-mentioned parameters with an aim to describe the merit
and demerit of the proposed schemes in comparison to a few existing
schemes.

We use the first definition for QE which was proposed by Cabello in
his seminal paper \citep{C2000} i.e., QE is defined as $\eta_{1}=\frac{b_{s}}{q_{t}+b_{t}}$,
here $b_{s}$ is the expected number of secret classical bits distributed
among the legitimate parties, $q_{t}$ is the number of qubits interchanged
via quantum channel during each step of the protocol, and $b_{t}$
is the classical bit information required to get the final key after
accomplishment of the protocol. Quantum resources are more expensive
and hard to maintain their coherence in comparison to communication
of classical information. Thus, we have also used another measure
that is frequently used for QE, i.e., $\eta_{2}=\frac{b_{s}}{q_{t}}$
\citep{TH+11}. We avoid accounting decoy state qubit to calculate
QE as decoy state is used to verify the channel security, not contribute
to generating the final key. To begin the comparison, we may consider
Huang et al.'s QKA protocol using quantum correlation in EPR pairs
with single qubit measurements to get the same key values between
two participants (Alice and Bob) \citep{HW+14}. This scheme is a
practically entangled version of BB84 QKD protocol with the only difference
in basis information is revealed before the measurement of Alice and
Bob. To perform this protocol one has to need quantum memory.
As Bob discloses the basis information to synchronize the measurement
outcomes for both the legitimate parties, there is a vulnerability
that allows eavesdropper to execute a photon number splitting attack. A similar
kind of approach was reported by Xu et al. using the GHZ state where
three parties get the same keys among themselves after the successful
execution of the protocol \citep{XW+14}. In that scheme, a three-qubit
quantum state is used as QR, and a subset of the initial sequence
is considered to establish the security which was prepared by the
party, $A_{1}$. To increase the QE, the legitimate parties measure
their single qubit systems only with computational basis ($B_{Z}$)
after channel security is established. The efficiency of their protocol
is $\eta_{1}=\frac{n-ns}{2n+ns}$\footnote{The values of the essential parameters for that protocol are: $b_{s}=n-ns$,
$q_{t}=2n$, $b_{t}=ns$, and the efficiency will be in the range $0.5\ge\eta\ge0$
with $s\in[0,1]$. The security will be increased as the value of
$s$ increases.}, here $s$ is the fraction of the qubits of the initial sequence
prepared by the first party ($A_{1}$) to verify the security of the
quantum channel. This scheme also needs quantum memory. In the same
year, Shukla et al. proposed a QKA scheme using EPR pairs, two-way
quantum channel \citep{SA+14} which is vulnerable to noise and quantum
memory. This scheme is also needed Pauli operation at Bob's end. That
scheme is essentially fabricated with little modification of the ping-pong
protocol \citep{BF02}. He et al. introduced a protocol
with enhanced efficiency ($\eta=0.5$), employing
a four-qubit cluster state which involves more QR. However, this approach
poses challenges in maintaining coherence of the four-qubit cluster
state and utilizes Pauli operators \citep{HM15}. Implementation of
this protocol requires a two-way quantum channel and quantum memory,
presenting a drawback in the context of current quantum communication
technology. Another
similar kind of scheme was proposed by Yang et al. using a four-qubit
cluster state, and cluster basis measurement \citep{YL+19}. The authors
used classical permutation operation in a complex manner to provide
more security from the adversary, Eve. This scheme inherently needs
quantum memory and Pauli operations\footnote{The efficiency for that scheme is $\eta_{1}=\frac{4n-nC}{4n+4n+nC}<0.5$
with $C\in[0,1]$, here $C$ is the checking set not act as decoy
state.}. 

Now, we explain the merit and demerit of our schemes for two-party
QKA with the previous ones. In our Protocol 2, we use EPR pair and
single-photon states which is convenient with the present technology.
There is no need for quantum memory to perform our scheme, and the
one-way quantum channel is used to avoid unnecessary channel noise,
unlike previously discussed protocols. Our protocol also not required
Pauli operations and has efficiencies $\eta_{1}=0.33$ and $\eta_{2}=1$.
Some situation demands the presence of a dishonest/untrusted third
party for performing a QKA protocol between two legitimate parties;
our Protocol 1 satisfies that demand too. In that context, Tang et
al. proposed the CQKA protocol using GHZ state, Hadamard gate, and
Pauli operation \citep{TS+20}. To perform their protocol they use
two-way quantum channel between Alice and Bob and quantum memory which
are the major demerit of their protocol. In our proposed
protocol, Protocol 1 (CQKA), we only require an EPR pair and CNOT
operation to implement our scheme using current technology. This is
especially advantageous when communication involves an untrusted third
party facilitating QKA between two legitimate parties. Notably, our
scheme does not necessitate the use of quantum memory and two-way
quantum channel, distinguishing it from the other protocols. The analysis
of the above discussion is summarized in the Table \ref{tab:Explicit-comparison}.

\begin{table}
\begin{centering}
\begin{tabular}{|c|c|c|c|c|c|c|c|}
\hline 
Protocol & NoP & QR & QC & QM & TR & QE $(\eta_{1})$ & QE $(\eta_{2})$\tabularnewline
\hline 
Huang et al. \citep{HW+14} & 2 & EPR pair & one-way & Y & N & $\frac{n}{2n}=0.5$ & $\frac{n}{n}=1$\tabularnewline
\hline 
Xu et al. \citep{XW+14} & 3 & GHZ state & one-way & Y & N & $\frac{n-ns}{2n+ns}<0.5$ & $\frac{n-s}{2n}<0.5$\tabularnewline
\hline 
Shukla et al. \citep{SA+14} & 2 & EPR pair & two-way & Y & N & $\frac{n}{2n+n}=0.33$ & $\frac{n}{2n}=0.5$\tabularnewline
\hline 
He et al. \citep{HM15} & 2 & four-qubit cluster state  & two-way & Y & N & $\frac{4n}{4n+4n}=0.5$ & $\frac{4n}{4n}=1$\tabularnewline
\hline 
Yang et al. \citep{YL+19} & 2 & four-qubit cluster state & one-way & Y & N & $\frac{4n-nC}{4n+4n+nC}<0.5$ & $\frac{4n-C}{4n}<1$\tabularnewline
\hline 
Tang et al. \citep{TS+20} & 2 & GHZ state & two-way & Y & Y & $\frac{2n}{6n+n}=0.285$ & $\frac{2n}{6n}=0.33$\tabularnewline
\hline 
Our Protocol 1 & 2 & EPR pair, single qubit  & one-way & N & Y & $\frac{n}{2n+3n}=0.2$ & $\frac{n}{2n}=0.5$\tabularnewline
\hline 
Our Protocol 2 & 2 & EPR pair, single qubit state & one-way & N & N & $\frac{n}{n+2n}=0.33$ & $\frac{n}{n}=1$\tabularnewline
\hline 
\end{tabular}
\par\end{centering}
\caption{Explicit comparison with previous protocols. Y - required, N - not
required, QR - quantum resources, QC - quantum channel, QM - quantum
memory, TR - third party, QE - quantum efficiency, NoP - number of
parties.\label{tab:Explicit-comparison}}
\end{table}

\section{Discussion\protect\label{sec:Discussion}}

Important conditions that a protocol for QKA needs to satisfy are
discussed in Section \ref{sec:Introduction}. Subsequently, in Section
\ref{sec:New-QKA-protocol}, we have shown that the schemes proposed
here satisfy the \emph{correctness} condition. Here, we briefly discuss
the \emph{fairness} condition using a specific scenario for our protocol
which is described in Tables \ref{tab:Relation of final key and announcement for P1}
and \ref{tab:Relation of final key and announcement for P2}.
As we have already mentioned that the \emph{fairness }condition demands
that an individual party cannot manipulate or influence the final
agreement key alone. If we consider for our protocol that Alice (Bob)
wants to influence the final key, i.e., her (his) choice of qubits
in sequence $S_{{\rm A}}(S_{{\rm B}})$ are all
in state $|0\rangle$ or $|1\rangle$, nevertheless, she (he) cannot
introduce bias in the final agreement key. One can easily justify
that demand by taking a specific example from the Table \ref{tab:Relation of final key and announcement for P1}
and Eqs. (\ref{eq:=00005CPsi2})-(\ref{eq:=00005CPsi8}). Without
loss of generality, we suppose that Alice chooses all the $|0\rangle$
states, and Bob prepares states in his qubit sequence randomly\footnote{Charlie's state is also random between $|\phi^{+}\rangle$ and $|\phi^{-}\rangle$
and corresponding bit values are $0$ and $1$, respectively.}. Then they (Alice and Bob) perform the CNOT operation on their qubits
with the qubits which Charlie sends to them. From Eqs (\ref{eq:=00005CPsi2})-(\ref{eq:=00005CPsi8}),
we obtain the final identical agreement keys in a manner that $0$
and $1$ appear with equal probability. Further, the \emph{fairness
}condition needs to satisfy the equal contribution of the legitimate
parties on the final agreement key. In the Step 4 of Protocol 1, Alice
and Bob independently prepare $n$ single qubit sequences $S_{{\rm A}}$
and $S_{{\rm B}}$ in $Z$ basis with a random choice of state $|0\rangle$
or $|1\rangle$, and in Step 5 they perform the CNOT operation. After
that, Alice and Bob do the Bell measurement on their two-qubit system
and then decide on the final key with the help of their measurement
results and the announcements (in Step 7 and Step 8) which is elaborately
mentioned in Table \ref{tab:Relation of final key and announcement for P1}
and \ref{tab:Relation of final key and announcement for P2}.
Hence, we can also see the contributions of the two parties are equal
in the final key. So, we can conclude that our protocol is satisfied
the \emph{fairness} condition. Another important condition is \emph{security}
which is also elaborately addressed in Section \ref{sec:Security-Analysis}.

\begin{figure}
\begin{centering}
\includegraphics[scale=0.5]{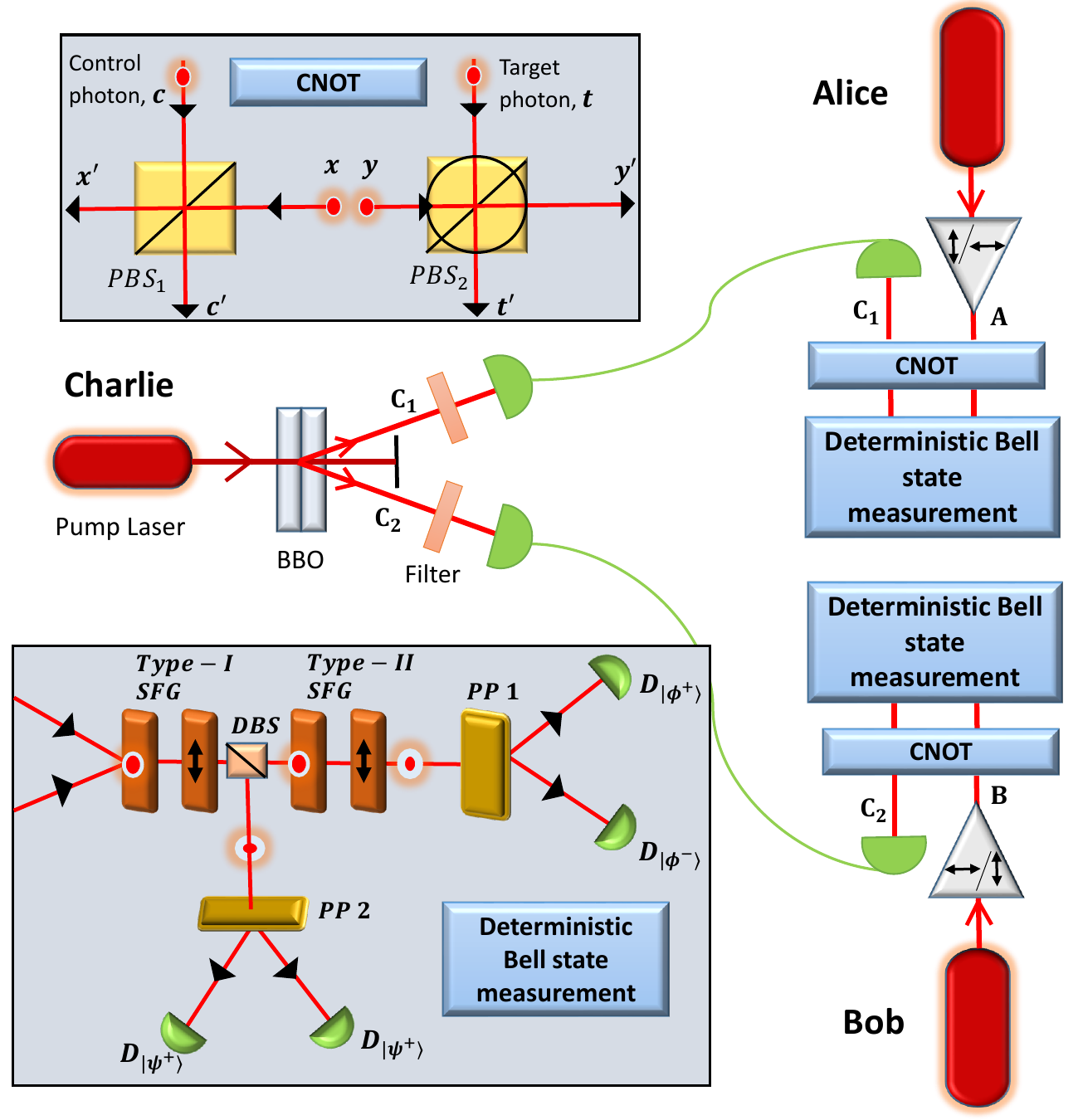}
\par\end{centering}
\caption{\protect\label{fig:Experimental_Setup}The optical
design of the CQKA scheme incorporates entangled photons, a CNOT operation,
and a comprehensive Bell state measurement. Entangled photons (Bell
states) can be generated using two thick $\beta$-barium borate (BBO)
nonlinear crystals arranged in a cross-configuration \citep{PPH+22}.
Bell state measurement is performed through nonlinear interactions,
specifically sum frequency generation (SFG) of type-I and type-II.
A dichroic beam splitter (DBS) is utilized, along with two $45^{\circ}$
projectors, labeled PP1 and PP2 \citep{SKP21}. The polarizing beam
splitter $({\rm PBS}_{1})$, ${\rm PBS}_{2}$ is configured with $45^{\circ}$
orientation by integrating a half-wave plate (HWP) into each input
and output of a standard polarizing beam splitter \citep{ZZC+05}.}
\end{figure}

Recently, numerous experimental implementations of quantum communication have
been conducted using both free-space links \citep{LCP+22,SJG+21}
and fiber-based links \citep{BRM+20}. In the context of quantum networks,
wires representing trajectories facilitate the transmission of quantum
systems \citep{RRE+21}. For such networks, global phase tracking
and strong reference signals are critical, and significant efforts
have been made to address these challenges efficiently \citep{ZLX+23}.
Experiments involving single-photon quantum communication represent
an essential step toward advancing future technologies \citep{CBD+23}.
Here, we propose an experiment to implement our scheme. The experimental
setup to implement this quantum communication scheme involves three
primary components (see Figure \ref{fig:Experimental_Setup}). Charlie\textquoteright s
entangled photon source, Alice and Bob\textquoteright s state preparation
stations, and their Bell-state measurement apparatus. Charlie generates
a Bell state $\left(|\phi^{\pm}\rangle=\frac{1}{\sqrt{2}}\left(|00\rangle\pm|11\rangle\right)\right)$
using a nonlinear crystal in a spontaneous parametric down-conversion
(SPDC) process \citep{PPH+22}, with one photon sent to Alice and
the other to Bob through fiber-optic or free-space links equipped
with polarization stabilization. Alice and Bob independently prepare
single-photon states in the $Z$ basis ($|0\rangle$ or $|1\rangle$)
using heralded single-photon sources and polarization control elements
such as waveplate(s) and polarizer(s). At each station, Alice and
Bob implement a linear-optics-based CNOT gate \citep{ZZC+05}, where
Charlie\textquoteright s entangled photon acts as the control qubit
and their locally prepared photon is the target qubit. The CNOT gate
is realized using polarizing beam splitter (${\rm PBS}_{1}$ and ${\rm PBS}_{2}$),
waveplate, and postselection techniques. The Bell-state measurement
is performed deterministically using nonlinear interactions through
sum frequency generation (SFG) of type-I and type-II processes \citep{SKP21}.
A DBS separates the generated sum-frequency photons, while two $45^{\circ}$
polarization projectors (PP1 and PP2) analyze the output states. This
setup ensures deterministic Bell-state identification with high fidelity.
High-efficiency single-photon detectors and a synchronized timing
system are employed for accurate detection and data collection, enabling
the successful implementation of the protocol and verification of
entanglement correlations.

\section{Conclusion\protect\label{sec:Conclusion}}

In some cases of key agreement between two parties may require a presence
of an untrusted third party. The quantum counterpart for such scenarios
requires QKA protocol with a controller untrusted third party.
Our protocol is designed in such a way that the requirements of the
controller in a QKA protocol is satisfied. In Section \ref{sec:New-QKA-protocol}
we elaborately present our protocol with the basic ideas in a step-wise
manner. There we explain how the quantum behavior of a qubit can
be exploited to design such kind of protocol(s) to fulfill the requirement
of practical applications. However, there is a certain requirement
to successfully execute a QKA protocol which is mentioned in Section
\ref{sec:Introduction}. The required security proof against various
attack scenarios is presented in Section \ref{sec:Security-Analysis}.
Here, we explain some important aspects of security analysis. To design
our Protocol 1, one needs to have Bell state which is prepared by
an untrusted third party, Charlie, and distributed to the legitimate
parties. After having the first and second qubits of the Bell state
prepared by Charlie, Alice and Bob perform the CNOT operation according
to our protocol. At this point, it is clear that, if Eve wants to
perform an intercept and resend attack on the channel, she can predict
only the Bell state which was prepared by Charlie as these channel
qubits are no longer correlated by Alice and Bob's operations for
that time being. This Bell state information $(k_{C})$ is also announced
by Charlie at the end of the protocol. So no advantage will be obtained
by Eve by performing such an attack. We also consider the attack when
Eve is trying to impersonate the third party, named\emph{
impersonated fraudulent attack}. Our protocol is
secure against this attack which is also analytically proved. One
more and most important security analysis is done against the collective
attack. This analysis proves our protocols are secure against more
generalized attack scenarios too\footnote{Our Protocol 2 is a special case of Protocol 1. Hence,
the security analysis for Protocol 1 is also applicable to Protocol
2.}. We also establish the probability of successfully
obtaining the final two-bit key by Eve is very less (can be eliminated)
with $n=6$ (see Figure \ref{fig:P vs n and d}).
Further, our security analysis may be extended by considering the
inner product of Eve's ancillary state as a complex quantity, and
each ancilla system can be expressed with a complete orthogonal basis
on the Hilbert space $\mathcal{\mathscr{H}^{\mathrm{\mathscr{E}}}}$
\citep{LF+11}. We also estimate the tolerable error bound for our
scheme by considering that approach and determine the tolerable QBER
as the function of $\alpha$. We analyze our scheme under various
noisy channel conditions and demonstrate the fidelity variation as
a function of the channel parameter. Our results show that the fidelity
does not decrease significantly, even for higher values of the noisy
channel parameter. In Section \ref{sec:Comparison}, we also discuss
the merit and demerit of our protocol with the same class of protocols
that were proposed in the recent past. Our protocol is shown to be
more feasible to implement with the available present technology and
more efficient in comparison with many of the recent schemes which
we discussed earlier (see Section \ref{sec:Comparison}) with a new
demanding feature of the presence of a third party.

\subsection*{Acknowledgment: }

Authors acknowledge the support from Interdisciplinary Cyber Physical
Systems (ICPS) programme of the Department of Science and Technology
(DST), India, Grant No.: DST/ICPS/QuST/Theme-1/2019/6 (Q46). AD
also thanks Sandeep Mishra and Prof. Subhashish Banerjee for some useful discussion.

\subsection*{Author contributions: }

Both the authors contributed equally to conceptualizing the problem. AD performed most of the calculations and prepared the first draft of the manuscript. AP supervised the work, validated the results, and edited the manuscript.

\subsection*{Consent to participate: }

Informed consent was obtained from all authors.

\subsection*{Data availability}

Data sharing is not applicable to this article as no datasets were
generated or analyzed during the current study.

\subsection*{Competing interests}

The authors declare that they have no competing interests.

\bibliographystyle{apsrev}
\bibliography{newQKA}

\end{document}